\documentclass[11pt]{article}
\usepackage{amssymb,amsmath,epsfig}
\usepackage{xcolor}
\usepackage{ulem}

\def\one{{{{\rm 1} \kern -.19em {\rm l}}}}
\def\C{{{{\rm {\mbox{\small l}}} \kern -.50em {\rm C}}}}
\def\R{{{{\rm l} \kern -.15em {\rm R}}}}
\def\N{{{{\rm l} \kern -.15em {\rm N}}}}
\def\E{{{{\rm l} \kern -.15em {\rm E}}}}
\def\P{{{{\rm l} \kern -.15em {\rm P}}}}
\def\Z{{{{\rm Z} \kern -.35em {\rm Z}}}}
\def\1{{{{\rm 1} \kern -.35em {\rm 1}}}}

\begin{document}
\begin{sloppypar}
\vspace*{0cm}
\begin{center}
{\setlength{\baselineskip}{1.0cm}{ {\large{\bf
DIRAC SYSTEMS WITH MAGNETIC FIELD AND POSITION-DEPENDENT MASS: DARBOUX TRANSFORMATIONS AND 
EQUIVALENCE WITH GENERALIZED DIRAC OSCILLATORS\\}} }}
\vspace*{1.0cm}
{\large{\sc{Axel Schulze-Halberg}}}$^\dagger$ ~~and~~ {\large{\sc{Pinaki Roy}}}$^{\ddagger, \ast}$
\end{center}
\noindent \\
$\dagger~$Department of Mathematics and Actuarial Science and Department of Physics, Indiana University Northwest, 3400 Broadway,
Gary IN 46408, USA, E-mail: axgeschu@iun.edu \\ \\
$\ddagger~$Atomic Molecular and Optical Physics Research Group, Advanced Institute of Materials Science, 
Ton Duc Thang University, Ho Chi Minh City, Vietnam \\ \\
$\ast~$Faculty of Applied Sciences, Ton Duc Thang University, Ho Chi Minh City, Vietnam, E-mail: pinaki.roy@tdtu.edu.vn
\vspace*{.3cm}
\begin{abstract}
\noindent
We construct a Darboux transformation for a class of two-dimensional Dirac systems at zero energy. Our starting 
equation features a position-dependent mass, a matrix potential, and an additional degree of freedom that 
can be interpreted either as a magnetic field perpendicular to the plane or a generalized Dirac oscillator interaction. 
We obtain a number of Darboux-transformed Dirac equations for which the zero energy solutions are exactly known.  

\end{abstract}


\section{Introduction}
Ever since the experimental realization of graphene \cite{geim} there has been a rising interest in Dirac materials and 
their applications. The distinguishing feature of Dirac materials such as graphene is that 
low-energy charge carriers behave like relativistic massless particles. As such, their dynamics within a monolayer of the 
material can be described through the two-dimensional, massless Dirac equation. We point out that this is not 
true anymore if several layers are present, such as in bilayer graphene \cite{mccann}. There is a vast amount of literature on 
Dirac materials and their applications, such that we refer the reader to the comprehensive reviews \cite{cayssol} \cite{wehling} 
and references therein. One of the standing tasks in the field is the confinement of charge carriers within a Dirac material, 
where the effect of Klein tunneling \cite{klein} has to be overcome. An overview of the problem and resolutions that 
have been proposed is given in \cite{downingsum} \cite{downingzero}. As pointed out in the latter references, 
a variety of methods has been explored for confining relativistic particles in Dirac materials, including the 
introduction of a position-dependent mass, and coupling the system to magnetic fields. Both of these 
generalizations have been implemented in Dirac systems. For example, Dirac systems with magnetic fields were studied 
on a hyperbolic graphene surface \cite{kizi}, under the presence of nonuniform fields \cite{downingmag}, 
within the minimal-length context \cite{menculini}, among others. Position-dependent masses were used in 
determining scattering states \cite{chabab}, systems with spatially variable Fermi velocity \cite{downing} \cite{gr} and 
generalized Dirac oscillators \cite{ho1}. Such oscillators, initially introduced as systems linear in momentum and 
coordinate variables \cite{mosh}, are closely related to Dirac models coupled to magnetic fields. Applications include 
experimental realizations of Dirac oscillators \cite{villafane}, their coupling to electric fields \cite{laba}, and within a 
rotating reference frame \cite{strange}. Interestingly, it has been shown that in $(2+1)$ dimensions the Dirac oscillator 
is equivalent to  a spin $1/2$ particle in a magnetic field \cite{castro}. Independent of the particular Dirac system that is studied regarding charge 
carrier confinement, its exactly-solvable particular cases play an important role. One of the most effective methods for 
finding and generating such rare cases is the Darboux transformation, also frequently known as 
supersymmetric quantum mechanics (SUSY) or intertwining technique \cite{cooper}. While upon introduction it applied to linear, 
second-order equations only \cite{darboux} \cite{moutard1} \cite{moutard2}, in the meantime the formalism of 
Darboux transformations has been adapted to a wide variety of systems governed by linear and nonlinear equations, 
including matrix differential equations like the Dirac equation. Comprehensive reviews of Darboux transformations can be found in \cite{gu} and 
\cite{matveev}. As far as Dirac systems are concerned, recent applications of the Darboux transformation include 
the case of magnetic fields, see for example \cite{castillo} \cite{astorga} \cite{midya}. The purpose of the present work is to 
generate solvable cases of the two-dimensional Dirac equation with position-dependent mass function $m$ and 
coupling to a magnetic field and a scalar potential. Effectively, our approach will generate a wide variety of cases, as 
the system considered here is equivalent to a generalized Dirac oscillator model or to an inhomogeneous magnetic field. 
From the application point of view, the $m=0$ scenario may be used to describe motion of 
electrons in gapless graphene in the presence of electromagnetic fields while $m\neq0$ scenario may used for 
gapped graphene \cite{peres} \cite{costa}. Let us now briefly discuss the 
method we will be using for generating solvable cases of the Dirac model. While the standard Darboux transformation 
has been extensively applied to the Dirac equation \cite{castillo} \cite{astorga} \cite{midya} \cite{xbatekat} \cite{xbatroy}, 
in the present work we apply a different Darboux transformation that was introduced in 
\cite{ustinov} \cite{lin} and later reformulated in \cite{xbathighx}. This Darboux transformation applies to a specific type 
of Schr\"odinger-like equation that can be obtained by suitably decoupling the Dirac equation. 
After application of the Darboux transformation we match the resulting Schr\"odinger-type equation to a 
form that can be put back into Dirac form. 
The remainder of this work is organized as follows: section 2 presents a brief review of the Darboux transformation for 
Schr\"odinger-type equations we will be using here. In section 3 we construct the generalization of the Darboux transformation 
to our Dirac scenario, while section 4 is devoted to examples. In section \ref{matrix} we shall consider the same 
system as in earlier sections except that matrix scalar potentials will be considered. Finally, section \ref{con} is 
devoted to a conclusion.

\section{Preliminaries}
Let us first summarize results from \cite{xbathighx}. The starting point is the following pair of Schr\"odinger-type equations
\begin{eqnarray}
\psi_0''(x)-\left[\epsilon^2+\epsilon~X_0(x)+Y_0(x) \right]~\psi_0(x) &=& 0 \label{lssei} \\[1ex]
\psi_n''(x)-\left[\epsilon^2+\epsilon~X_n(x)+Y_n(x) \right]~\psi_n(x) &=& 0, \label{lsset} 
\end{eqnarray}
where the prime denotes differentiation, $\epsilon$ is a real-valued constant, the functions $X_j$, $Y_j$, $j=0,n$, 
are sufficiently smooth and independent of $\epsilon$, and $\psi_0$, $\psi_n$ represent the respective 
solutions for a natural number $n$. In resemblance to the conventional Schr\"odinger equation we will refer to 
$\epsilon$ as energy and to $X_j$, $Y_j$, $j=0,n$, as potential terms. We will now define a Darboux transformation 
that interrelates the two equations (\ref{lssei}) and (\ref{lsset}). To this end, we define functions $v_j$, $j=0,...,n-1$, 
through
\begin{eqnarray}
v_j(x) &=& \exp\left[(\epsilon-\lambda_j)~x \right] h_j(x),~~~j=0,...,n-1, \label{vj}
\end{eqnarray}
where $h_j$, $j=0,...,n-1$, are solutions of the initial equation (\ref{lssei}) at energies $\lambda_j$, $j=0,...,n-1$, 
respectively, such that the constants $\lambda_0$, 
$\lambda_1$,..., $\lambda_{n-1},~\epsilon$ are pairwise different. The solutions $h_j$, $j=0,...,n-1$, are called 
transformation functions. We are now ready to define our $n$-th order Darboux transformation. This 
transformation $\psi_n$ of the solution $\psi_0$ to (\ref{lssei}) is given by
\begin{eqnarray}
\psi_n(x) &=& \frac{W_{v_{n-1},\psi_0}(x)}{\sqrt{\hat{W}_{v_{n-1}}(x)~W_{v_{n-1}}(x)}}. 
\label{main1}
\end{eqnarray}
Here, the quantities $W_{v_{n-1}}$ and $W_{n-1,\psi_0}$ stand for the Wronskians of $v_0,...,~v_{n-1}$ and of 
$v_0,...,~v_{n-1},~\psi_0$, respectively. Furthermore, $\hat{W}_{v_{n-1}}$ is given by 
\begin{eqnarray}
\hat{W}_{v_{n-1}}(x) &=& \left(-2 \right)^n \frac{1}{G(x)}~W_{v_0,...,v_{n-1},F}(x),~~~\mbox{where}~
G(x) ~=~ \exp\left[\frac{1}{2}~\int\limits^x V_0(t)+2~\epsilon~dt \right]. \label{F}
\end{eqnarray}
The function $\psi_n$ solves our transformed Schr\"odinger-type equation (\ref{lsset}) if the potential terms meet 
the following constraints
\begin{eqnarray}
X_n(x) &=& X_0(x)+ \frac{d}{dx}~\log\left[\frac{\hat{W}_{v_{n-1}}(x)}{W_{v_{n-1}}(x)} \right] \label{main2} \\[1ex]
Y_n(x) &=& Y_0(x) -\frac{n}{2}~X_0'(x)+\frac{X_0(x)}{2} \left\{ \frac{d}{dx}~\log\left[\frac{\hat{W}_{v_{n-1}}(x)}{W_{v_{n-1}}(x)} 
\right] \right\}
+
\frac{3~[\hat{W}'_{v_{n-1}}(x)]^2}{4~\hat{W}_{v_{n-1}}(x)^2}+ \nonumber \\[1ex]
&+&\frac{3~[W_{v_{n-1}}'(x)]^2}{4~W_{v_{n-1}}(x)^2}- \frac{\hat{W}_{v_{n-1}}'(x)~W_{v_{n-1}}'(x)}{2~\hat{W}_{v_{n-1}}(x)~W_{v_{n-1}}(x)}
-\frac{\hat{W}_{v_{n-1}}''(x)}{2~\hat{W}_{v_{n-1}}(x)}-\frac{W_{v_{n-1}}''(x)}{2~W_{v_{n-1}}(x)}. \label{main3}
\end{eqnarray}
In summary, the quantities (\ref{main1}), (\ref{main2}), and (\ref{main3}) determine the interrelations between the initial 
(\ref{lssei}) and the transformed Schr\"odinger-type equation (\ref{lsset}) and their corresponding solutions.

\section{Darboux transformations for the Dirac equation}
The purpose of this section is to construct a Darboux transformation for the two-dimensional Dirac equation at 
zero energy. The principal idea used for our construction 
is to connect an initial and transformed Dirac equation with Schr\"odinger-type counterparts of the form 
(\ref{lssei}) and (\ref{lsset}), respectively.

\paragraph{Decoupling the initial Dirac equation.} 
We start out from the initial Dirac equation in the form
\begin{eqnarray}
\left\{\sigma_x \left[p_x-i~\sigma_z~f(x) \right]+\sigma_y~p_y+\sigma_z~m(x)+V(x)~I_2 \right\} \Psi(x,y) &=& 0, \label{dirac}
\end{eqnarray}
where $\sigma_x,~\sigma_y,~\sigma_z$ are the usual Pauli matrices, $p_x,~p_y$ denotes the momentum 
operators, and $\Psi$ is the two-component solution. Furthermore, $f$ can be interpreted as a generalized oscillator term, 
$m$ denotes the position-dependent mass, and $V~I_2$ represents a scalar potential function $V$ multiplied by the 
$2 \times 2$ identity matrix. It is interesting to note that equation (\ref{dirac}) may also be written as
\begin{eqnarray} 
\left\{\sigma_x~ p_x+\sigma_y~\left[p_y-f(x)\right]+\sigma_z~m(x)+V(x)~I_2 \right\} \Psi(x,y) &=& 0. \label{diracmag}
\end{eqnarray}
In this form, our Dirac equation describes a particle that is subjected to a magnetic field \cite{castro}: our function $f$ can be interpreted as a component of the vector potential $A$, given by
\begin{eqnarray}
A(x) &=& \left(0,-f(x),0 \right)^T. \nonumber
\end{eqnarray}	
Consequently, the associated magnetic field $B$ is obtained by applying the curl. This yields
\begin{eqnarray}
B(x) &=& \left(0,0,-f'(x) \right)^T. \label{b}
\end{eqnarray}
For the following it does not make any difference if we consider our Dirac equation in the form (\ref{dirac}) or 
(\ref{diracmag}), as the only difference between the two forms is the 
interpretation of the function $f$. As an example let us mention that the massless case $m=0$ of 
our second form (\ref{diracmag}) describes a quasiparticle in graphene subjected to an inhomogeneous magnetic field 
perpendicular to the graphene sheet. Next, upon inserting the momentum operators and collecting terms, our equation (\ref{dirac}) 
can be written as follows
\begin{eqnarray}
-i~\frac{\partial \Psi(x,y)}{\partial x} -i~\frac{\partial \Psi(x,y)}{\partial y}+
\left(
\begin{array}{lll}
m(x) +V(x) & i~f(x) \\[1ex]
-i~f(x) & -m(x)+V(x)
\end{array}
\right)
\Psi(x,y) &=& 0. \label{dirac1}
\end{eqnarray}
Next, noting that the motion in $y-$direction is free, we introduce the solution components by setting
\begin{eqnarray}
\Psi(x,y) &=& \exp(i~k_y~y) \left(
\begin{array}{ll}
\Psi_1(x) \\[1ex]
\Psi_2(x)
\end{array}
\right), \label{psi}
\end{eqnarray}
where the real-valued constant $k_y$ denotes the momentum in the $y$-direction. Upon implementing (\ref{psi}) in our Dirac 
equation (\ref{dirac1}), the spinor components can be shown to follow the following pair of coupled equations
\begin{eqnarray}
-i~\Psi_2'(x)+\left[
-i~k_y+i~f(x)\right]\Psi_2(x)+
\left[ m(x)+V(x)\right]\Psi_1(x) &=& 0, \label{sys1} \\[1ex]
-i~\Psi_1'(x)+\left[
i~k_y-i~f(x)\right]\Psi_1(x)+
\left[- m(x)+V(x)\right]\Psi_2(x) &=& 0. \label{sys2}
\end{eqnarray}
In order to decouple this system, we solve the second equation with respect to $\Psi_2$. This gives
\begin{eqnarray}
\Psi_2(x) &=& \frac{[i~f(x)-i~k_y]~\Psi_1(x)+i~\Psi_1'(x)}{V(x)-m(x)}. \label{psi2}
\end{eqnarray}
We substitute this setting into the first equation (\ref{sys1}), along with the definition
\begin{eqnarray}
\Psi_1(x) &=& \sqrt{m(x)-V(x)}~\psi_0(x), \label{psi1}
\end{eqnarray}
introducing a function $\psi_0$. This renders our equation (\ref{sys1}) in the following form
\begin{eqnarray}
\psi_0''(x)- \left[k_y^2+k_y~X_0(x)+Y_0(x)\right] \psi_0(x) &=& 0, \label{sse}
\end{eqnarray}
where the functions $X_0$ and $Y_0$ are given by
\begin{eqnarray}
X_0(x) &=& -2~f(x) - \frac{m'(x)-V'(x)}{m(x)-V(x)} \label{x0} \\[1ex]
Y_0(x) &=& \frac{1}{4~[m(x)-V(x)]^2} ~
\Bigg\{4~f(x)^2[m(x)-V(x)]^2+4~f(x)~[m(x)-V(x)] \times \nonumber \\[1ex]
&\times& [m'(x)-V'(x)]+3~[m'(x)-V'(x)]^2+2~[m(x)-V(x)]
\big\{2~[m(x)-V(x)] \times \nonumber \\[1ex]
&\times&[m(x)^2-V(x)^2-f'(x)]-m''(x)+V''(x)\big\}
\Bigg\}. \label{y0}
\end{eqnarray}
We observe that the form of our equation (\ref{sse}) matches its general counterpart (\ref{lssei}) if we identify the 
parameters $\epsilon$ and $k_y$.

\paragraph{The transformed Dirac system.}
As a consequence of the matching we just completed, our Darboux transformation becomes applicable to (\ref{sse}). 
While the transformed solution (\ref{main1}) and its associated potential terms (\ref{main2}), (\ref{main3}) can be 
calculated in a straightforward manner, the remaining task is to use the latter results in order to set up a transformed 
Dirac equation of the type (\ref{dirac}). More precisely, this transformed Dirac equation reads
\begin{eqnarray}
\left\{\sigma_x \left[p_x-i~\sigma_z~\hat{f}(x) \right]+\sigma_y~p_y+\sigma_z~\hat{m}(x)+\hat{V}(x)~I_2 \right\} \hat{\Psi}(x,y) &=& 0, 
\label{diract}
\end{eqnarray}
where we must determine its transformed solution $\hat{\Psi}$, the term $\hat{f}$, the 
position-dependent mass $\hat{m}$, and the scalar potential function $\hat{V}$, multiplied by the $2 \times 2$ identity matrix. 
As in the case of our initial Dirac equation, we can rewrite its transformed counterpart (\ref{diract}) in the form
\begin{eqnarray}
\left\{\sigma_x ~p_x+\sigma_y~\left[p_y-{\hat f}(x)\right]+\sigma_z~\hat{m}(x)+\hat{V}(x)~I_2 \right\} \hat{\Psi}(x,y) = 0,
\label{diractmag}
\end{eqnarray}
which we understand to describe a system coupled to a magnetic field that is given by
\begin{eqnarray}
\hat{B}(x) &=& \left(0,0,-\hat{f}'(x) \right)^T. \label{bh}
\end{eqnarray}
Next, we will first find the latter three quantities, and 
afterwards construct the associated transformed Dirac solution. After applying the Darboux transformation (\ref{main1}) to 
equation (\ref{sse}), we obtain a transformed equation of the form (\ref{lsset}), that is
\begin{eqnarray}
\psi_n''(x)-\left[k_y^2+k_y~X_n(x)+Y_n(x) \right]~\psi_n(x) &=& 0, \label{sset} 
\end{eqnarray}
recall that $k_y$ replaces the parameter $\epsilon$ in (\ref{lsset}). The potential terms $X_n$ and $Y_n$ are given by 
(\ref{main2}) and (\ref{main3}), respectively, where $X_0$, $Y_0$ can be found in (\ref{x0}), (\ref{y0}). Hence, in the case of 
$X_n$ we have the following explicit form
\begin{eqnarray}
X_n(x) &=& -2~f(x) - \frac{m'(x)-V'(x)}{m(x)-V(x)} + \frac{d}{dx}~\log\left[\frac{\hat{W}_{v_{n-1}}(x)}{W_{v_{n-1}}(x)} \right]. 
\label{xn}
\end{eqnarray}
It is now important to understand that this expression must be cast in a shape resembling (\ref{x0}), such that it can be 
linked to our transformed Dirac scenario. This yields the condition
\begin{eqnarray}
-2~f(x) - \frac{m'(x)-V'(x)}{m(x)-V(x)} + \frac{d}{dx}~\log\left[\frac{\hat{W}_{v_{n-1}}(x)}{W_{v_{n-1}}(x)} \right] 
&=& -2~\hat{f}(x) - \frac{\hat{m}'(x)-\hat{V}'(x)}{\hat{m}(x)-\hat{V}(x)}. \label{xncon}
\end{eqnarray}
The same type of condition must hold for the second potential term $Y_n$. However, since the explicit form of 
this condition is very long, as it involves (\ref{main3}) and (\ref{y0}), we omit to state it here. Instead, we give it in 
abbreviated form as
\begin{eqnarray}
Y_n(x) &=& Y_0(x)_{\mid f \rightarrow \hat{f}, m \rightarrow \hat{m}, V \rightarrow \hat{V}}, \label{yncon}
\end{eqnarray}
The system of equations (\ref{xncon}), (\ref{yncon}) determines the above mentioned quantities that make up the transformed 
Dirac equation (\ref{diract}): the term $\hat{f}$, position-dependent mass $\hat{m}$, and scalar potential $\hat{V}$. We proceed by 
solving (\ref{xncon}) with respect to the transformed term $\hat{f}$. We obtain
\begin{eqnarray}
\hat{f}(x) &=& 2~f(x) + \frac{m'(x)-V'(x)}{m(x)-V(x)} -\frac{\hat{m}'(x)-\hat{V}'(x)}{\hat{m}(x)-\hat{V}(x)} - 
\frac{d}{dx}~\log\left[\frac{\hat{W}_{v_{n-1}}(x)}{W_{v_{n-1}}(x)} \right] \nonumber \\[1ex]
&=& 2~f(x) + \frac{d}{dx}~\log\left\{\frac{[m(x)-V(x)]~W_{v_{n-1}}(x)}{[\hat{m}(x)-\hat{V}(x)]~\hat{W}_{v_{n-1}}(x)} \right\}. 
\label{fh}
\end{eqnarray}
Note that we could have also solved (\ref{xncon}) with respect to the mass or the  potential, but this would have lead to an 
unsolvable second condition. For this reason, we go with our function (\ref{fh}). Substitution 
into the second condition (\ref{yncon}) and solving for the transformed scalar potential gives 
\begin{eqnarray}
\hat{V}(x) &=& \delta~\Bigg\{\frac{1}{4~[m(x)-V(x)]}~\Bigg[\hat{m}(x)^2+\frac{\Delta X_n'(x)}{2}+
\frac{\Delta X_n(x)~V'(x)}{2~[m(x)-V(x)]}-m(x)^2+V(x)^2 -\nonumber \\[1ex]
&-& -f(x)~\Delta X_n(x)+\frac{\Delta X_n(x)^2}{4}-\Delta Y_n(x)-
\frac{\Delta X_n(x)~m'(x)}{2~[m(x)-V(x)]}
\Bigg]
\Bigg\}^\frac{1}{2}, \label{vh}
\end{eqnarray}
where $\delta=\pm1$. For the sake of brevity we used the abbreviations
\begin{eqnarray}
\Delta X_n(x) ~=~ X_n(x)-X_0(x) \qquad \qquad \qquad \Delta Y_n(x) ~=~ Y_n(x)-Y_0(x), \label{deltas}
\end{eqnarray}
recall that the quantities involved here are defined in (\ref{main2}) and (\ref{main3}), respectively. 
Thus, we have now solved our system (\ref{xncon}), (\ref{yncon}) by determining the transformed 
function (\ref{fh}) and the transformed scalar potential (\ref{vh}). Note that the 
transformed position-dependent mass remains undetermined and can be set arbitrarily. It is important to 
point out that the transformed mass can always be chosen as zero, such that our transformed Dirac equation 
becomes massless. As mentioned above, this scenario particularly applies to charge carrier transport in Dirac 
materials like graphene.

\paragraph{The transformed Dirac solutions.}
It now remains to construct the solution of our transformed Dirac equation (\ref{diract}), which we will do in a way 
similar to its initial counterpart (\ref{psi}). We define the transformed solution in two-component form as
\begin{eqnarray}
\hat{\Psi}(x,y) &=& \exp(i~k_y~y) \left(
\begin{array}{ll}
\hat{\Psi}_1(x) \\[1ex]
\hat{\Psi}_2(x)
\end{array}
\right). \label{psih}
\end{eqnarray}
The component functions of this solution are interrelated by
\begin{eqnarray}
\hat{\Psi}_2(x) &=& \frac{[i~\hat{f}(x)-i~k_y]~\hat{\Psi}_1(x)+i~\hat{\Psi}_1'(x)}{\hat{V}(x)-\hat{m}(x)}, \label{psi2h}
\end{eqnarray}
note that this relation is in agreement with (\ref{psi2}). It remains to determine the first component in (\ref{psih}). 
To this end, let us compare the present case with the initial scenario, where the first component $\Psi_1$ of the Dirac solution 
(\ref{psi}) is linked to a solution $\psi_0$ of the Schr\"odinger equation (\ref{sse}) by means of (\ref{psi1}). This means 
that in the transformed scenario, the first solution component $\hat{\Psi}_1$ is related to the 
transformed Schr\"odinger solution $\psi_n$ as
\begin{eqnarray}
\hat{\Psi}_1(x) &=& \sqrt{\hat{m}(x)-\hat{V}(x)}~\psi_n(x), \label{psi1hpre}
\end{eqnarray}
recall that $\hat{V}$ is given in (\ref{vh}) and $\hat{m}$ is arbitrary. Next, we observe that the function 
$\psi_n$ entering in (\ref{psi1hpre}) is a solution of the transformed Schr\"odinger equation (\ref{lsset}). As such, it can 
be written using the Darboux transformation (\ref{main1}). This renders (\ref{psi1hpre}) in the form
\begin{eqnarray}
\hat{\Psi}_1(x) &=&  \sqrt{\hat{m}(x)-\hat{V}(x)}~
\frac{W_{v_{n-1},\psi_0}(x)}{\sqrt{\hat{W}_{v_{n-1}}(x)~W_{v_{n-1}}(x)}}. \label{psi1h}
\end{eqnarray}
Let us now establish the connection between the functions $v_j$, $j=0,...,n-1$ and our transformed Dirac 
equation (\ref{diract}). To this end, we take into account the definition (\ref{vj}) that introduces solutions 
$h_j$, $j=0,...,n-1$, of our initial Schr\"odinger equation (\ref{sse}). Upon using the same relation as in 
(\ref{psi1}), we find
\begin{eqnarray}
v_j(x) &=& \exp\left[(k_y-\lambda_j)~x \right] \sqrt{\frac{1}{m(x)-V(x)}}~\chi_j(x),~~~j=0,...,n-1, \label{chij}
\end{eqnarray}
where $\chi_j$ is the first component of a solution to our transformed Dirac equation (\ref{diract}) for $k_y=\lambda_j$, 
$j=0,...,n-1$. The associated second component can be found through the same transformation as used in (\ref{psi2}).

\section{Applications}
We will now present several applications for the Darboux transformation that was constructed in the previous section. While 
our construction's starting point is the initial Dirac equation (\ref{dirac}), from a practical point of view 
it is typically more efficient to use our Schr\"odinger-type equation (\ref{sse}) instead. The reason is that solutions of 
the latter equation can be found much more easily than of its Dirac counterpart. Once a solution to (\ref{sse}) 
is known, solutions, potentials, and terms for both our initial and transformed Dirac equation 
can be generated. We will follow this procedure in our subsequent examples. Due to the importance of the initial
Schr\"odinger-type equation (\ref{sse}) for the Darboux transformation we will now mention a particular simplification 
that arises when parameters are chosen suitably. The principal idea of this parameter choice is to remove the term 
proportional to $k_y$, that is, we impose the condition $X_0=0$ in (\ref{x0}). This condition can be fulfilled by 
choosing the term as
\begin{eqnarray}
f(x) &=& \frac{V'(x)-m'(x)}{2~[m(x)-V(x)]}. \label{fsim}
\end{eqnarray}
Upon substituting this into our equation (\ref{sse}), the remaining potential term (\ref{y0}) simplifies. We obtain 
\begin{eqnarray}
\psi_0''(x)+ \left[-k_y^2+V(x)^2-m(x)^2\right] \psi_0(x) &=& 0. \label{schr}
\end{eqnarray}
This equation can be interpreted as a conventional Schr\"odinger equation with energy $-k_y^2$ and potential 
$m^2-V^2$. Hence, we can choose the initial mass $m$ and potential $V$ in order to obtain a solvable Schr\"odinger 
equation (\ref{schr}). The only parameter restriction is that the energy must be negative. This is so because the 
parameter $k_y$ must be real-valued due to our definition (\ref{psi}) of the Dirac solution. Let us also point out that 
the term (\ref{fsim}) is determined once the mass $m$ and the potential $V$ have been chosen.

\subsection{First application}
\begin{figure}[h]
\begin{center}
\epsfig{file=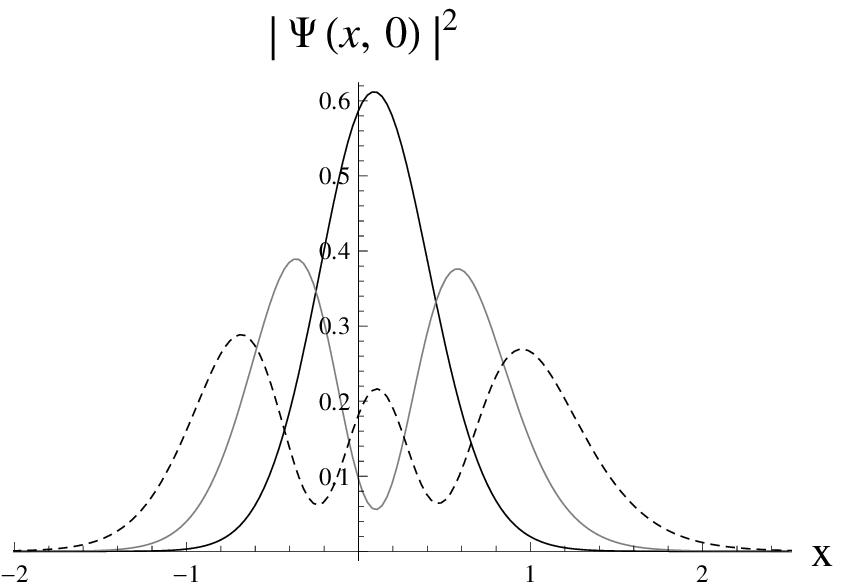,width=7.8cm}
\epsfig{file=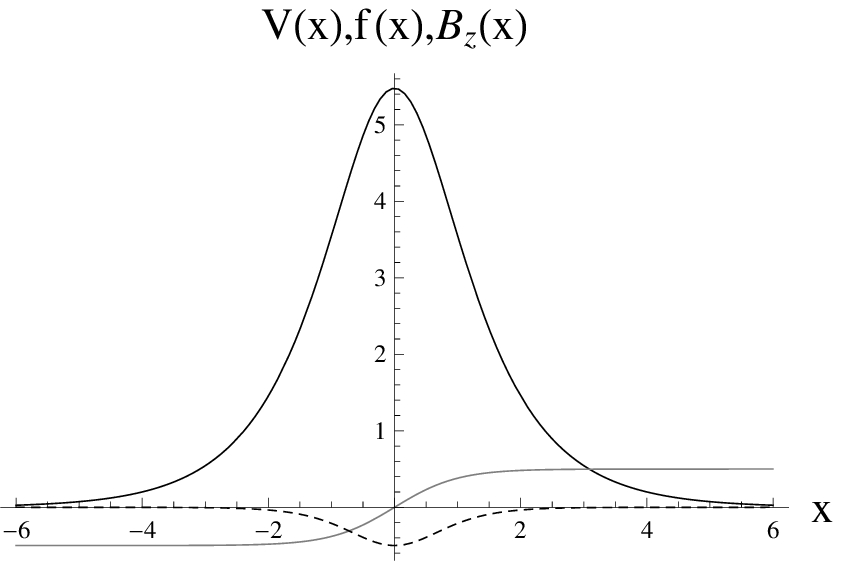,width=7.8cm}
\caption{Left plot: graphs of the normalized probability densities $|\Psi(x,0)|^2$ associated with the solution 
(\ref{psi}) for components (\ref{psi11}) and (\ref{psi21}). Parameter settings are $k_y=5$ (black solid curve), 
$k_y=4$ (gray curve), and $k_y=3$ (dashed curve). Right plot: the initial functions $V$ (black solid curve), 
$f$ (gray curve) from (\ref{set1}), and the $z-$component of the magnetic field (\ref{mag1}) for the mass $m=0$.
}
\label{fig1}
\end{center}
\end{figure}

Let us consider our initial Dirac equation (\ref{dirac}) or, equivalently, the form (\ref{diracmag}) for the 
following parameter settings
\begin{eqnarray}
f(x) ~=~ \frac{1}{2}~\tanh(x) \qquad \qquad V(x) ~=~ \sqrt{30}~\mbox{sech}(x) \qquad \qquad m(x) ~=~ 0. \label{set1}
\end{eqnarray}
Note that the factor $\sqrt{30}$ in the potential was chosen in order to obtain a certain amount of bound-state solutions 
to our Dirac equation, as will be demonstrated below. Observe further that the settings (\ref{set1}) render (\ref{dirac}) 
in massless form, such that it applies to Dirac materials like graphene. The functions from (\ref{set1}) 
are shown in the right part of figure \ref{fig1}. While $V$ stands for the scalar potential, the function $f$ can either denote a 
generalized oscillator term according to (\ref{dirac}) or it can represent a magnetic field within (\ref{diracmag}) that 
is found by means of (\ref{b}) as
\begin{eqnarray}
B(x) &=& \left(0,0,-\frac{1}{2}~\mbox{sech}(x)^2 \right)^T. \label{mag1}
\end{eqnarray}
Hence, the last component of the magnetic field has the shape of a pulse. We substitute our settings into the 
Schr\"odinger equation (\ref{sse}) that after simplification takes the form
\begin{eqnarray}
\psi_0''(x)- \left[k_y^2-30~\mbox{sech}(x)^2\right] \psi_0(x) &=& 0. \label{sse1}
\end{eqnarray}
We observe here that the term proportional to $k_y$ has vanished. This is so because our choice of parameters in (\ref{set1}) 
satisfies (\ref{fsim}). The general solution of equation (\ref{sse1}) can be written as
\begin{eqnarray}
\psi_{\scriptsize{\mbox{gen}}}(x) &=&c_1~ P_5^{k_y}\left[\tanh(x)\right]+c_2~Q_5^{k_y}\left[\tanh(x)\right], \label{psi0gen}
\end{eqnarray}
where $P$ and $Q$ stand for the associated Legendre function of the first and second kind, respectively \cite{abram}. 
In order to simplify calculations and to extract bound-state solutions, we will consider the following 
particular solution of equation (\ref{sse1}), obtained from the general case (\ref{psi0gen}) by setting $c_1=1$ and $c_2=0$
\begin{eqnarray}
\psi_0(x) &=& P_5^{k_y}\left[\tanh(x)\right], \label{psi01}
\end{eqnarray}
The function (\ref{psi01}) enables us to find a solution to 
our initial Dirac equation (\ref{dirac}) with the settings (\ref{set1}). Upon substitution of (\ref{psi01}) into 
(\ref{psi1}) and (\ref{psi2}), we obtain the component functions of the solution (\ref{psi}) as follows
\begin{eqnarray}
\Psi_1(x) &=& \sqrt{\mbox{sech}(x)}~P_5^{k_y}\left[\tanh(x)\right] \nonumber \\[1ex]
\Psi_2(x) &=& i~\sqrt{30}~
\sqrt{\mbox{sech}(x)} \times \label{psi11} \\[1ex]
& & \hspace{-1.5cm} \times~\Bigg\{
(k_y-6)~\cosh(x)~P_6^{k_y}\left[\tanh(x)\right] +[6~\sinh(x)-k_y~\cosh(x)]~P_5^{k_y}\left[\tanh(x)\right] \Bigg\}. 
\label{psi21}
\end{eqnarray}
The corresponding solution (\ref{psi}) of our initial Dirac equation (\ref{dirac}) represents bound states if the 
parameter $k_y$ attains integer values in the interval $[1,5]$. The left part of figure \ref{fig1} shows associated normalized 
probability densities.
 \noindent
We are now ready to apply our Darboux transformation.

\begin{figure}[h]
\begin{center}
\epsfig{file=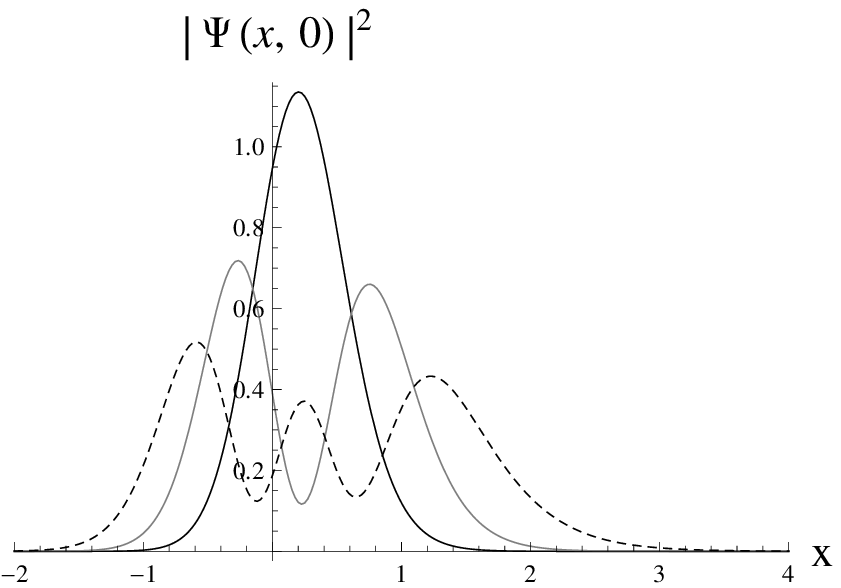,width=7.8cm}
\epsfig{file=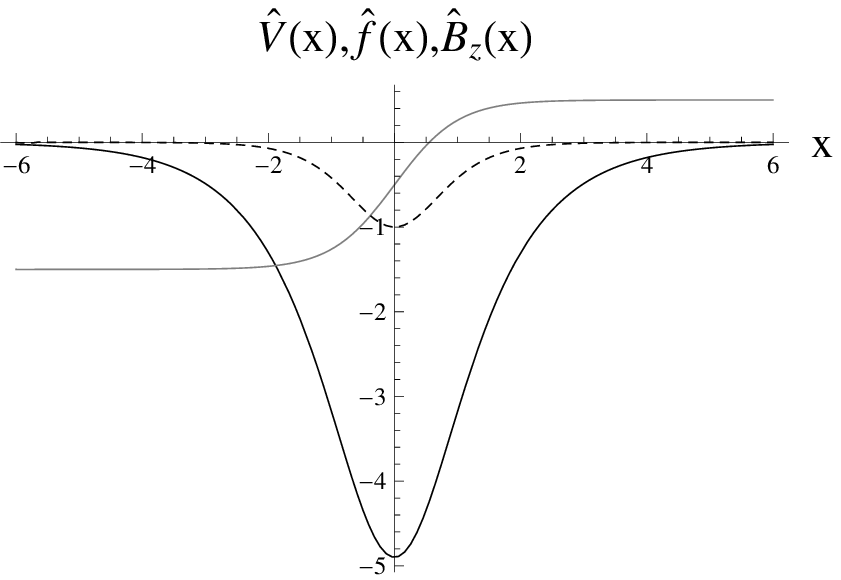,width=7.8cm}
\caption{Left plot: graphs of the normalized probability densities $|\hat{\Psi}(x,0)|^2$ associated with the solution 
(\ref{psih}) for components (\ref{psi11h0}) and (\ref{psi21h0}). Parameter settings are $\delta=-1$, $k_y=4$ (black solid curve), 
$k_y=3$ (gray curve), and $k_y=2$ (dashed curve). Right plot: the transformed functions $\hat{V}$ (black solid curve), 
$\hat{f}$ (gray curve) from (\ref{vh0fh0}), and the $z-$component of the associated magnetic field (\ref{bh}) (dashed curve)  
for $\delta=-1$ and the mass function (\ref{mh0}).}
\label{fig1_1}
\end{center}
\end{figure}

\begin{figure}[h]
\begin{center}
\epsfig{file=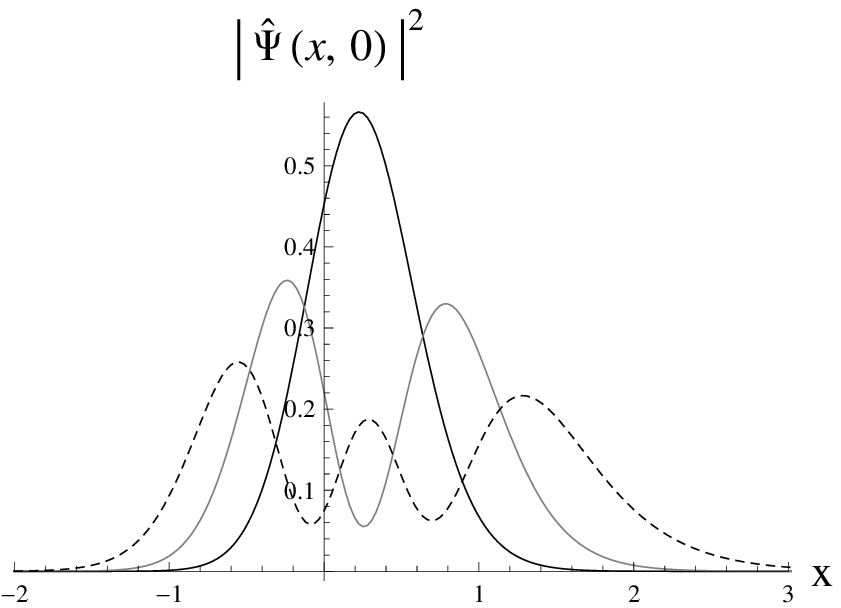,width=7.8cm}
\epsfig{file=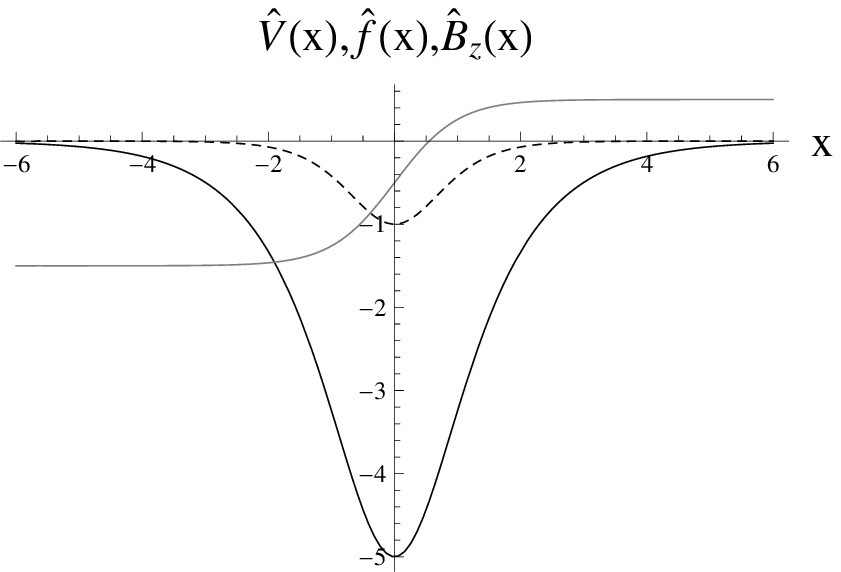,width=7.8cm}
\caption{Left plot: graphs of the normalized probability densities $|\hat{\Psi}(x,0)|^2$ associated with the solution 
(\ref{psih}) for components (\ref{psi11h}) and (\ref{psi21h}). Parameter settings are $\delta=-1$, $k_y=4$ (black solid curve), 
$k_y=3$ (gray curve), and $k_y=2$ (dashed curve). Right plot: the transformed functions $\hat{V}$ (black solid curve), 
$\hat{f}$ (gray curve) from (\ref{vh1fh1}), and the $z-$component of the associated magnetic field (\ref{bh}) (dashed curve) for $\delta=-1$ and the mass function (\ref{mh1}).}
\label{fig2}
\end{center}
\end{figure}

\begin{figure}[h]
\begin{center}
\epsfig{file=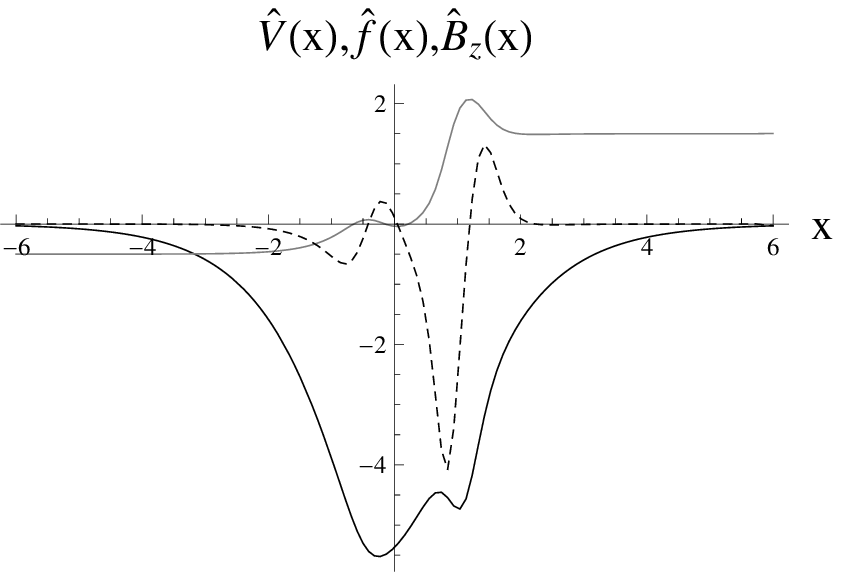,width=7.8cm}
\epsfig{file=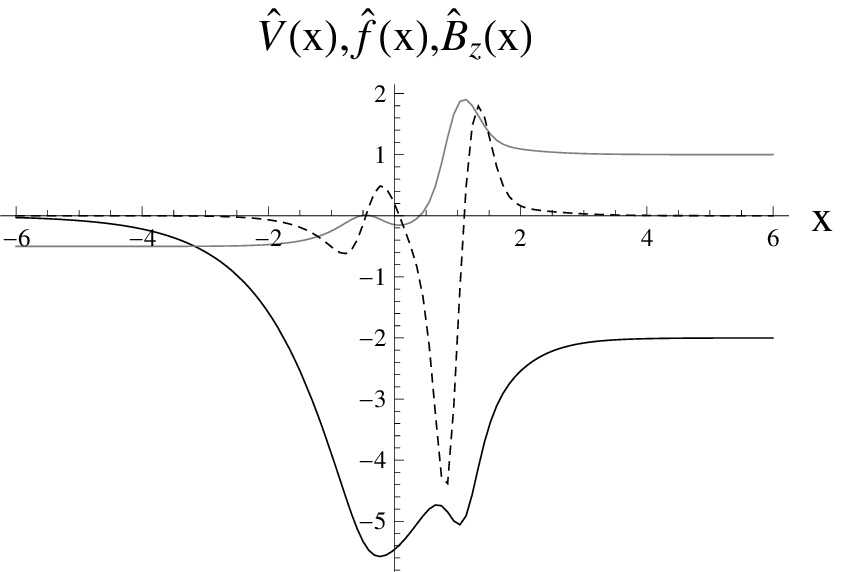,width=7.8cm}
\caption{Graphs of the transformed scalar potential (\ref{vh}) (black solid curves), the term (\ref{fh}) (gray curves), 
and the $z-$component of the associated magnetic field (\ref{bh}) (dashed curves) 
for the mass functions $\hat{m}(x) =0$ (left plot) and 
$\hat{m}(x) =1+\tanh(x)$ (right plot) and the settings (\ref{set1}), (\ref{h01a}).}
\label{fig3}
\end{center}
\end{figure}

\paragraph{First-order Darboux transformation.} Let us first perform a transformation of order one by 
setting $n=1$ throughout (\ref{main1}), (\ref{main2}), and (\ref{main3}). We choose the transformation function 
$h_0$ from (\ref{psi01}) for the transformation parameter $\lambda_0=5$, that is, we set
\begin{eqnarray}
h_0(x) &=& \psi_0(x)_{|k_y=5} ~=~ P_5^{5}\left[\tanh(x)\right] ~=~ -945 \left[1-\tanh(x)^2\right]^\frac{5}{2}. \label{h01}
\end{eqnarray}
We substitute this function into (\ref{vj}) and the Darboux transformation (\ref{main1}), (\ref{main2}), (\ref{main3}), 
and we afterwards plug the results along with our settings (\ref{set1}) into the transformed scalar 
potential (\ref{vh}) and the term (\ref{fh}). This gives for the choice $\delta=-1$
\begin{eqnarray}
\hat{V}(x) &=& -\sqrt{\hat{m}(x)^2+24~\mbox{sech}(x)^2} \label{vh1} \nonumber \\[1ex]
\hat{f}(x) &=& -\frac{1}{2}+\frac{1}{2}~\tanh(x)-\frac{\hat{m}'(x)-\hat{V}'(x)}{2~\hat{m}(x)-2~\hat{V}(x)}, \label{fh1}
\end{eqnarray}
We observe that the these function is defined on the whole real line, 
provided the mass fuction $\hat{m}$ is real-valued and nonnegative. The associated solution of our 
transformed Dirac equation (\ref{diract}) is obtained through the formulas (\ref{psi2h}) and (\ref{psi1h}). We do not state 
the corresponding general expressions in explicit form due to their length. Instead, we give examples for specific 
mass functions. In our first example we consider the massless scenario, that is, we set
\begin{eqnarray}
\hat{m}(x) &=& 0. \label{mh0}
\end{eqnarray}
This choice renders our scalar potential (\ref{vh}) and the function (\ref{fh}) in the form
\begin{eqnarray}
\hat{V}(x) &=& -2~\sqrt{6}~\mbox{sech}(x) \qquad \qquad \qquad \hat{f}(x)~=~\tanh(x)-\frac{1}{2}, \label{vh0fh0}
\end{eqnarray}
where we set $\delta =-1$. Graphs of these functions are shown in the right part of figure \ref{fig1_1}. In the form 
(\ref{diract}) of our Dirac equation, our function $\hat{f}$ stands for a generalized oscillator term, while in the 
equivalent form (\ref{diractmag}) we use (\ref{bh}) to determine the magnetic field that is represented by $\hat{f}$. 
We obtain
\begin{eqnarray}
\hat{B}(x) &=& \left(0,0,-\mbox{sech}(x)^2\right)^T. \nonumber
\end{eqnarray}
It remains to construct a solution of our transformed Dirac equation. To this end, we will now 
use (\ref{vh0fh0}) to evaluate the components (\ref{psi2h}) and (\ref{psi1h}) of our transformed Dirac solution 
(\ref{psih}). This gives us
\begin{eqnarray}
\hat{\Psi}_1(x) &=& \frac{1}{\sqrt{10}~\cosh(x)^\frac{3}{2}~\sqrt{[-1-\tanh(x)]}} \times \nonumber \\[1ex]
& & \hspace{-1.9cm}
\times~\Bigg\{
(k_y-6)~\cosh(x)~P_6^{k_y}\left[\tanh(x)\right]+[5-k_y+11~\cosh(x)+11~\sinh(x)]~P_5^{k_y}\left[\tanh(x)\right]
\Bigg\} \label{psi11h0} \\[1ex]
\hat{\Psi}_2(x) &=&-
i~\frac{1}{8~\sqrt{15}~\cosh(x)^\frac{3}{2}~\sqrt{-1-\tanh(x)}} \times \nonumber \\[1ex]
&\times&
\Bigg\{-2~(k_y-6)~\cosh(x)[(7-k_y)~\cosh(x)~P_7^{k_y}\left[\tanh(x)\right]+
[2~(k_y-2)~\cosh(x)- \nonumber \\[1ex]
&-&
19~\sinh(x)]~P_6^{k_y}\left[\tanh(x)\right]+
\{[-60-4~k_y+k_y^2+(72-4~k_y+k_y^2)~\cosh(2x)-\nonumber \\[1ex]
&-&
6~(3~k_y-4)~\sinh(2x)]
\}
~P_5^{k_y}\left[\tanh(x)\right]
\Bigg\}.
\label{psi21h0}
\end{eqnarray}
Normalized probability densities associated with these solutions are shown in the left part of figure \ref{fig1_1}. We 
observe that the solutions are of bound-state type if $k_y=1,2,3,4$. In other words, the momentum 
$k_y$  can not take arbitrary values and must necessarily be quantized in order for bound states to exist.
 \noindent
Let us now switch to a massive case of our Dirac equation (\ref{dirac}) by choosing
\begin{eqnarray}
\hat{m}(x) &=& \mbox{sech}(x). \label{mh1}
\end{eqnarray}
Upon plugging this mass into the transformed scalar potential (\ref{vh}) and our function (\ref{fh}), the latter quantities 
are rendered in the form
\begin{eqnarray}
\hat{V}(x) &=& -5~\mbox{sech}(x) \qquad \qquad \qquad \hat{f}(x)~=~\tanh(x)-\frac{1}{2}, \label{vh1fh1}
\end{eqnarray}
where we set $\delta=-1$. The solution components (\ref{psi2h}) and (\ref{psi1h}) evaluate as follows
\begin{eqnarray}
\hat{\Psi}_1(x) &=& \sqrt{\frac{1}{10~\cosh(x)~[-1-\tanh(x)]}} \times \nonumber \\[1ex]
&\times&
\Bigg\{
(k_y-6)~P_6^{k_y}\left[\tanh(x)\right]+[5-k_y+11~\tanh(x)]~P_5^{k_y}\left[\tanh(x)\right]
\Bigg\} \label{psi11h} \\[1ex]
\hat{\Psi}_2(x) &=& i~
\frac{1}{-12~\sqrt{10}~\cosh(x)^\frac{3}{2}~\sqrt{-1-\tanh(x)}} \times \nonumber \\[1ex]
&\times&
\Bigg\{
2~(k_y-6)~\cosh(x)[(k_y-7)~\cosh(x)~P_7^{k_y}\left[\tanh(x)\right]+
[-2~(k_y-2)~\cosh(x)+ \nonumber \\[1ex]
&+&
19~\sinh(x)]~P_6^{k_y}\left[\tanh(x)\right]+
\{(k_y-10)(k_y+6)+[72+(k_y-4)~k_y]~\cosh(2x)+\nonumber \\[1ex]
&+&
6~(-3~k_y+4)~\sinh(2x)]\}~P_5^{k_y}\left[\tanh(x)\right]
\Bigg\}. \label{psi21h}
\end{eqnarray}
The associated solution (\ref{psih}) represents bound states if $k_y$ takes integer values in the interval $[1,4]$, as 
we can observe in the left part of figure \ref{fig2}.
 \noindent
Next, we repeat the application of our first-order Darboux transformation, where we switch out our transformation function 
(\ref{h01}) as follows
\begin{eqnarray}
h_0(x) &=& Q_5^{5.51}\left[\tanh(x)\right]. \label{h01a}
\end{eqnarray}
Note that we obtained this transformation function from the general solution (\ref{psi0gen}) by setting 
$c_1=0$ and $c_2=1$. Upon performing the Darboux transformation 
(\ref{main1}), (\ref{main2}), (\ref{main3}) for the settings (\ref{set1}), the transformation function (\ref{h01a}), and the 
two masses $\hat{m}(x) =0$ and $\hat{m}(x) =1+\tanh(x)$, we 
obtain the results shown in figure \ref{fig3}.
 \noindent
For the sake of brevity we do not include the explicit expressions for the quantities shown in figure \ref{fig3}, as they can 
be obtained in a straightforward manner by plugging the chosen mass function into (\ref{fh}) and (\ref{vh}).

\paragraph{Generalization and bound states.} We will now generalize the previous example by introducing a nonzero initial position-dependent 
mass function. Our new settings that replace (\ref{set1}) are given by
\begin{eqnarray}
f(x) ~=~ \frac{1}{2}~\tanh(x) \qquad \qquad V(x) ~=~ \sqrt{30}~\mbox{sech}(x) \qquad \qquad 
m(x) ~=~ \alpha~\sqrt{30}~\mbox{sech}(x). \label{set2}
\end{eqnarray}
Here, $\alpha$ is a real-valued parameter that controls the strength of the mass function. We observe that the latter 
function is proportional to the scalar potential. We will comment on this property below in a more general context. 
The purpose of the present example is to study the effect of $\alpha$ on the transformed system, in particular on the 
discrete spectrum. To this end, let us substitute the settings (\ref{set2}) into the Schr\"odinger equation (\ref{sse}). 
We obtain
\begin{eqnarray}
\psi_0''(x)- \left[k_y^2-30~\mbox{sech}(x)^2~(1-\alpha^2)\right] \psi_0(x) &=& 0. \label{sse2}
\end{eqnarray}
From the Schr\"odinger perspective, the potential associated with this equation has the form of a single-well, 
the depth of which is determined by $\alpha$. If $\alpha$ vanishes, the well has maximum depth, such that 
the system supports five bound states \cite{cooper}. As the value of $\alpha$ increases, the potential well's 
depth decreases, as well as the number of supported bound states. When $\alpha=1$, the potential 
vanishes and no bound states are supported by the system. This can be verified by looking at the actual 
bound-state solutions of (\ref{sse2}). Their general form reads
\begin{eqnarray}
\psi_0(x) &=& P_{-\frac{1}{2}+\frac{1}{2} \sqrt{121-120 \alpha^2}}^{k_y}\left[
\tanh(x)\right], \label{psi02}
\end{eqnarray}
where the lower index of the Legendre function must be a positive integer, and the upper index must be an 
integer. In addition, (\ref{psi02}) must satisfy the condition
\begin{eqnarray}
-\frac{1}{2}+\frac{1}{2}~ \sqrt{121-120 ~\alpha^2}-|k_y| &=& N, ~~~N=0,1,2,..., \label{alpha}
\end{eqnarray}
where $|k_y|$ can take integer values in the interval $[1,5]$. For any given value of $k_y$, 
the number of solutions to equation (\ref{alpha}) decreases as $\alpha$ raises. The values of $\alpha$ 
that generate a specific number of supported bound states is shown in table 1.
\begin{table}[h]
\begin{center}
\vspace{.5cm}
\begin{tabular}[c]{|l|l|} 
\hline 
Number of bound states & $\alpha$ \\[1ex] 
\hline 
5 & 0 \\[1ex]
\hline  
4 & $\sqrt{\frac{1}{3}}$ \\[1ex]
\hline
3 & $\sqrt{\frac{3}{5}}$ \\[1ex]
\hline  
2 & $\sqrt{\frac{2}{5}}$ \\[1ex]
\hline
1 & $\sqrt{\frac{14}{15}}$ \\[1ex]
\hline
\end{tabular}
\caption{Number of bound states and associated parameter value $\alpha$.} \nonumber
\end{center}
\end{table}
It is straightforward to verify that the behavior of the bound-state solutions to (\ref{sse2}) is the same for our 
initial Dirac equation. In particular, the values of $\alpha$ given in table 1 remain valid for the initial Dirac case 
(\ref{dirac}). For the sake of brevity we omit to show the actual solution. 
As far as the transformed Dirac equation is concerned, the numbers from table 1 are not valid anymore because the 
number of supported bound states depends not only on $\alpha$, but also on the transformation function used in the 
Darboux transformation, and on the transformed position-dependent mass function. The only general statement that 
can be made is that the number of supported bound states decreases if $\alpha$ increases.

\begin{figure}[h]
\begin{center}
\epsfig{file=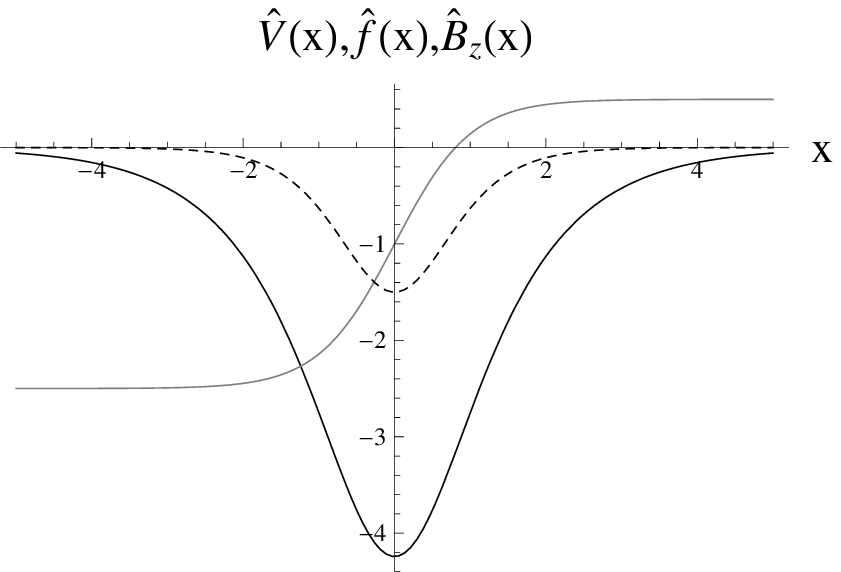,width=7.8cm}
\epsfig{file=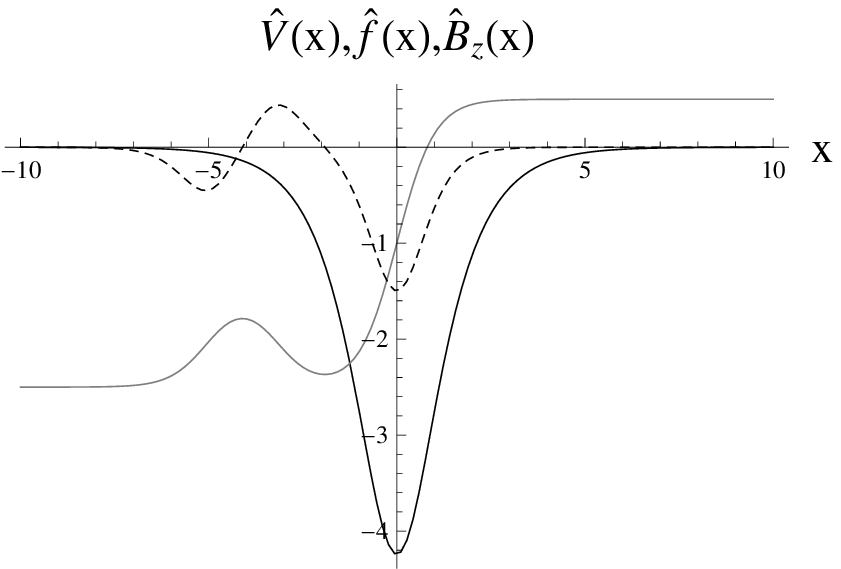,width=7.8cm}
\caption{Graphs of the transformed functions (\ref{vh311}) (black solid curve), (\ref{fh311}) (gray curve), 
and the $z-$component of the associated magnetic field (\ref{bh}) (dashed curve) 
for the mass functions $\hat{m}(x)=0$ (left plot) and 
$\hat{m}(x)=\mbox{sech}(x+5)$ (right plot).}
\label{fig31}
\end{center}
\end{figure}

\begin{figure}[h]
\begin{center}
\epsfig{file=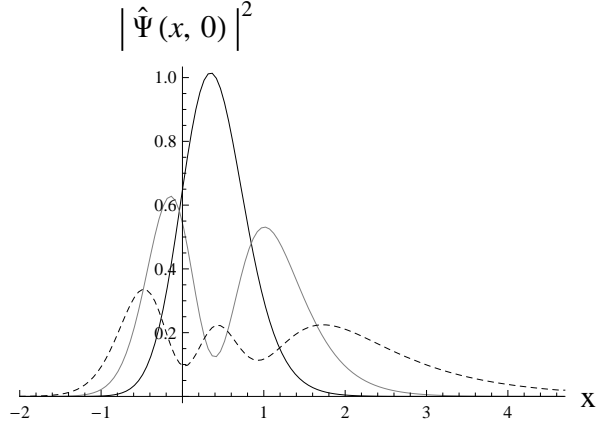,width=7.8cm}
\caption{Graphs of the normalized probability densities $|\hat{\Psi}(x,0)|^2$ associated with the solution 
(\ref{psih}) of our transformed Dirac equation (\ref{diract}) for $\hat{m}=0$ and 
the settings (\ref{vh311}), (\ref{fh311}). 
Parameter values are $\delta=-1$, $k_y=3$ (black solid curve), 
$k_y=2$ (gray curve), and $k_y=1$ (dashed curve).}
\label{fig32}
\end{center}
\end{figure}

\paragraph{Second-order Darboux transformations.} Let us return to our Dirac equation (\ref{dirac}) with 
the settings (\ref{set1}), and perform a Darboux transformation of second order. 
This requires two transformation function $h_0$ and $h_1$ that we define as 
\begin{eqnarray}
h_0(x) &=& \psi_0(x)_{|k_y=5} ~=~ P_5^{5}\left[\tanh(x)\right] ~=~ -945 \left[1-\tanh(x)^2\right]^\frac{5}{2} \nonumber \\[1ex]
h_1(x) &=& \psi_0(x)_{|k_y=4} ~=~ P_5^{4}\left[\tanh(x)\right] ~=~ 945~\tanh(x) \left[1-\tanh(x)^2\right]^2. \label{h11}
\end{eqnarray}
Note that $h_0$ is the same as its counterpart in (\ref{h01}). We now apply our Darboux transformation by 
substituting (\ref{h01}), (\ref{h11}) into (\ref{vj}) and 
(\ref{main1}), (\ref{main2}), (\ref{main3}) for $n=2$. The results in combination with our settings (\ref{set1}) determine 
the transformed scalar potential (\ref{vh}) and the function (\ref{fh}). We find
\begin{eqnarray}
\hat{V}(x) &=& - \sqrt{\hat{m}(x)+18~\mbox{sech}(x)^2} \label{vh311} \\[1ex]
\hat{f}(x) &=& -1+\tanh(x)-\frac{\hat{m}'(x)-\hat{V}'(x)}{2~\hat{m}(x)-2~\hat{V}(x)}, \label{fh311}
\end{eqnarray}
recall that we set $\delta=-1$. Graphs of these two functions are shown in \ref{fig31} for specific masses. While the 
left part of the latter figure displays the massless scenario, in the right part we create a deformation of the 
graphs around the point $x=-5$ by introducing a mass that has the shape of a pulse around that point.
\noindent
Within the interpretation of our Dirac equation in the form (\ref{diractmag}), the function (\ref{fh311}) generates a magnetic 
field that is found by means of (\ref{bh}). For the case of vanishing mass $\hat{m}=0$, the latter magnetic field reads
\begin{eqnarray}
\hat{B}(x) &=& \left(0,0,-3~\sqrt{2}~\mbox{sech}(x)^2 \right)^T. \nonumber
\end{eqnarray}
The solutions of our transformed Dirac equation (\ref{diract}) associated with the quantities (\ref{vh311}), 
(\ref{fh311}) are shown in figure \ref{fig32} as normalized probability densities.
 \noindent

\subsection{Second application}
In this section we will present another example of applying our Darboux transformation to the initial Dirac equation 
in any of the equivalent forms (\ref{dirac}) or (\ref{diracmag}). We will choose the following parameter setting for the 
initial scenario
\begin{eqnarray}
f(x) ~=~0 \qquad \qquad V(x) ~=~ \alpha~\mbox{sech}(x) \qquad \qquad m(x) ~=~ 0, \label{set3}
\end{eqnarray}
where $\alpha$ is a negative real number. Upon implementation of these settings, our initial Dirac equation 
renders in massless form with the scalar potential $V$, special cases of which are shown in the right part of figure \ref{fig3}. 
Now, insertion of the settings (\ref{set3}) into our Schr\"odinger equation (\ref{sse}) 
renders the latter in the form
\begin{eqnarray}
\psi_0''(x)-\left[k_y^2+k_y~\tanh(x)+\left(\frac{1}{4}-\alpha^2 \right) \mbox{sech}(x)^2+\frac{1}{4}
\right] \psi_0(x) &=& 0. \label{sse3}
\end{eqnarray}
We observe that in comparison to its counterpart (\ref{sse1}), this equation contains a term proportional to $k_y$ because 
the settings (\ref{set3}) do not comply with the condition (\ref{fsim}). Equation (\ref{sse3}) is exactly-solvable with 
particular solution
\begin{eqnarray}
\psi_0(x) &=& \cosh(x) \left[1-\tanh(x)\right]^{\frac{1}{2}+k_y} \left[-1+\tanh(x)\right]^{\frac{1}{4}-\frac{k_y}{2}} 
\left[1+\tanh(x)\right]^{\frac{1}{4}+\frac{k_y}{2}} \times \nonumber \\[1ex]
&\times& {}_2F_1\Bigg[
\frac{1}{2}+k_y-q,\frac{1}{2}+k_y+q,\frac{3}{2}+k_y,\frac{1}{1+\exp(2 x)}
\Bigg], \label{psi03}
\end{eqnarray}
where ${}_2F_1$ stands for the hypergeometric function \cite{abram}. Before we focus on our Darboux transformation, 
let us construct a solution of our initial Dirac equation (\ref{dirac}). To this end, we substitute (\ref{psi03}) 
into the components (\ref{psi1}) and (\ref{psi2}) of (\ref{psi}). We obtain the result
\begin{eqnarray}
\Psi_1(x) &=& \sqrt{\mbox{sech}(x)}~\psi_0(x) \label{psi13} \\[1ex]
\Psi_2(x) &=& -\frac{i}{2~\alpha}~\sqrt{\mbox{sech}(x)}~\Bigg\{\Bigg[2~k_y~\cosh(x)+\sinh(x)\Bigg]~\psi_0(x)-2~\cosh(x)~\psi_0'(x)\Bigg\}, \label{psi23}
\end{eqnarray}
where the function $\psi_0$ is defined in (\ref{psi03}). The components (\ref{psi13}), (\ref{psi23}) represent bound states 
if $\alpha$ and $k_y$ are interrelated as 
\begin{eqnarray}
\frac{1}{2}+k_y+\alpha &=& -N, ~~~N=0,1,2,3,...\nonumber
\end{eqnarray}
We observe that this is precisely the condition under which the first argument of the hypergeometric function in 
(\ref{psi03}) turns into a nonpositive integer. As a result, the latter function degenerates to a polynomial. The left 
part of figure \ref{fig3} visualizes an example for a specific parameter setting.
\begin{figure}[h]
\begin{center}
\epsfig{file=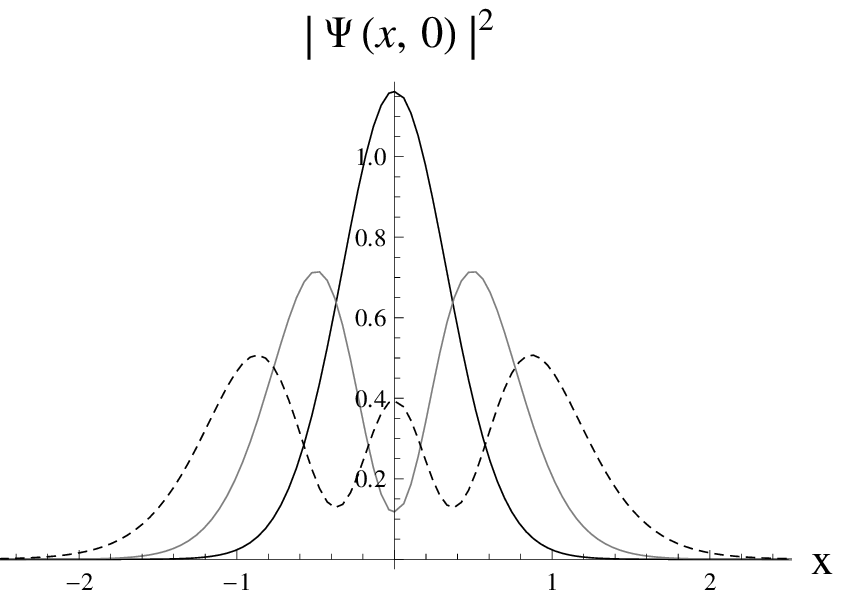,width=7.8cm}
\epsfig{file=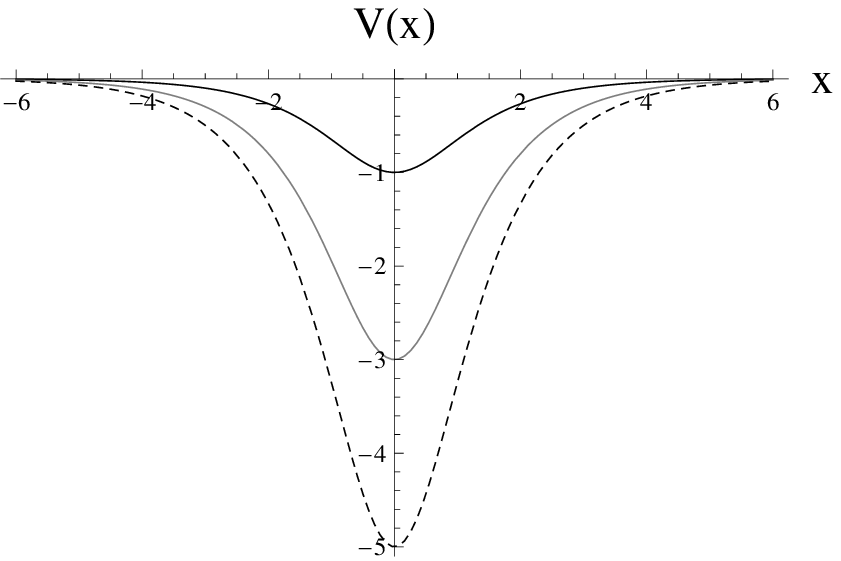,width=7.8cm}
\caption{Left plot: graphs of the normalized probability densities $|\Psi(x,0)|^2$ associated with the solution 
(\ref{psi}) for components (\ref{psi13}) and (\ref{psi23}). Parameter settings are $\alpha=-5$, $k_y=9/2$ (black solid curve), 
$k_y=7/2$ (gray curve), and $k_y=5/2$ (dashed curve). Right plot: the initial scalar potential in (\ref{set3}) for the parameter 
settings $\alpha=-1$ (black solid curve), $\alpha=-3$ (gray curve), and $\alpha=-5$ (dashed curve).}
\label{fig4}
\end{center}
\end{figure} \noindent
\paragraph{First-order Darboux transformation.} In order to keep calculations simple, we restrict ourselves 
to the case $\alpha=-1$ in (\ref{set3}). We choose our transformation function from (\ref{psi03}) as
\begin{eqnarray}
h_0(x) &=& \psi_0(x)_{|k_y=-1} ~=~ \frac{\exp\left(\frac{3 x}{2}\right)}{2~\sqrt{1+\exp(2 x)}}, \label{h03}
\end{eqnarray}
where for the sake of simplicity we switched from hyperbolic to exponential functions. In the next step we 
plug (\ref{h03}) into (\ref{vj}) and into the Darboux transformation (\ref{main1}), (\ref{main2}), (\ref{main3}) 
for $n=1$. Afterwards we insert the results in combination with our settings (\ref{set3}) into the function (\ref{fh}) and 
the scalar potential (\ref{vh}). We obtain
\begin{eqnarray}
\hat{V}(x) &=&- \sqrt{
\hat{m}(x)^2+\frac{12~\exp(2 x)}{\left[3+\exp(2x)\right]^2}} \label{vh3} \\[1ex]
\hat{f}(x) &=& -\frac{1}{2}+\frac{3}{3+\exp(2x)}-\frac{\hat{m}'(x)-\hat{V}'(x)}{2~\hat{m}(x)-2~\hat{V}(x)}. \label{fh3}
%\hat{f}(x) &=& -\frac{\left[-3+\exp(2 x)\right] \hat{m}(x)+\left[3+\exp(2 x)\right] \hat{m}'(x)}
%{2~\sqrt{12~\exp(2 x)+\left[3+\exp(2 x)\right]^2 \hat{m}(x)^2}}. \label{fh3}
\end{eqnarray}
As in the previous occurrences we have set $\delta=-1$. If the mass $\hat{m}$ is regular on the whole 
real line, so are the two functions (\ref{vh3}) and (\ref{fh3}) because the denominators are nonnegative. Figure \ref{fig5} shows 
graphs of the transformed quantities $\hat{V}$ and $\hat{f}$ for two particular mass choices. We observe that the 
first of these choices $\hat{m}=0$ makes the term (\ref{fh3}) vanish.
\begin{figure}[h]
\begin{center}
\epsfig{file=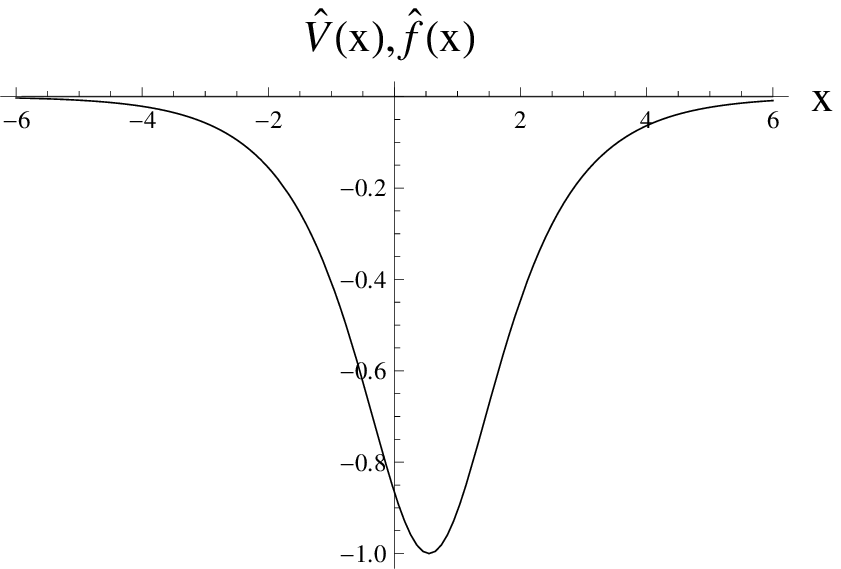,width=7.8cm}
\epsfig{file=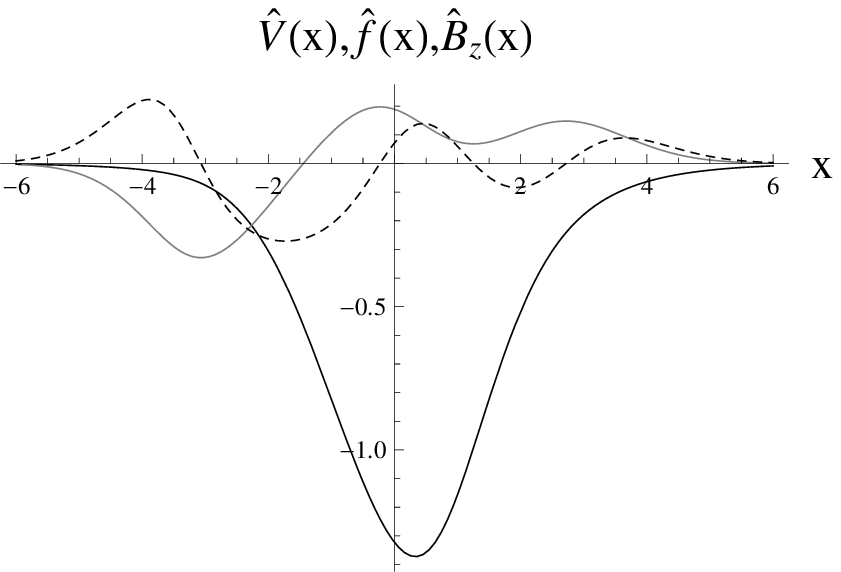,width=7.8cm}
\caption{Graphs of the transformed scalar potential (\ref{vh3}) (black solid curves), the transformed generalized 
oscillator term (\ref{fh3}) (gray curve), and the $z-$component of the associated magnetic field (\ref{bh}) 
(dashed curve) for the mass functions $\hat{m}(x) =0$ (left plot) and 
$\hat{m}(x) =\exp(-x^2/3)$ (right plot) and the settings (\ref{set3}), (\ref{h03}), $\delta=-1$.}
\label{fig5}
\end{center}
\end{figure} \noindent
We omit to show explicit expressions of the associated solutions to our transformed Dirac equation.

\begin{figure}[h]
\begin{center}
\epsfig{file=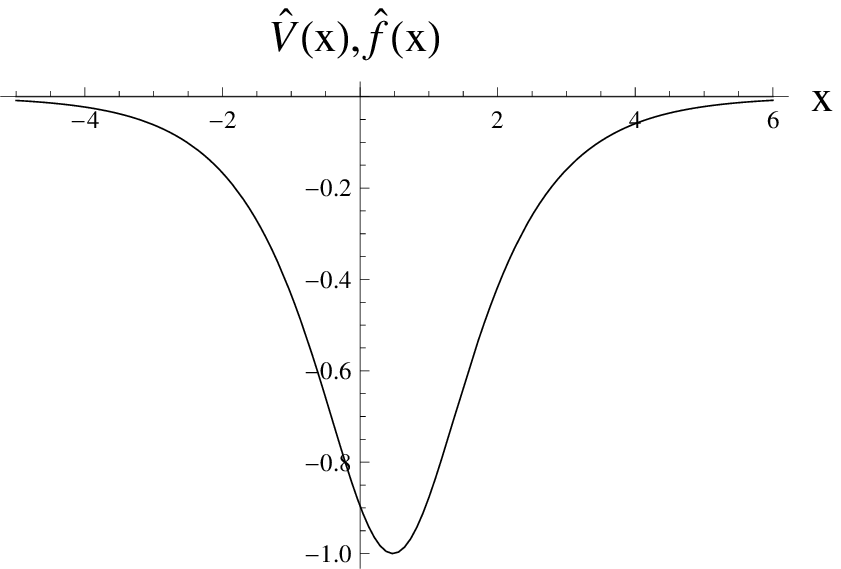,width=7.8cm}
\epsfig{file=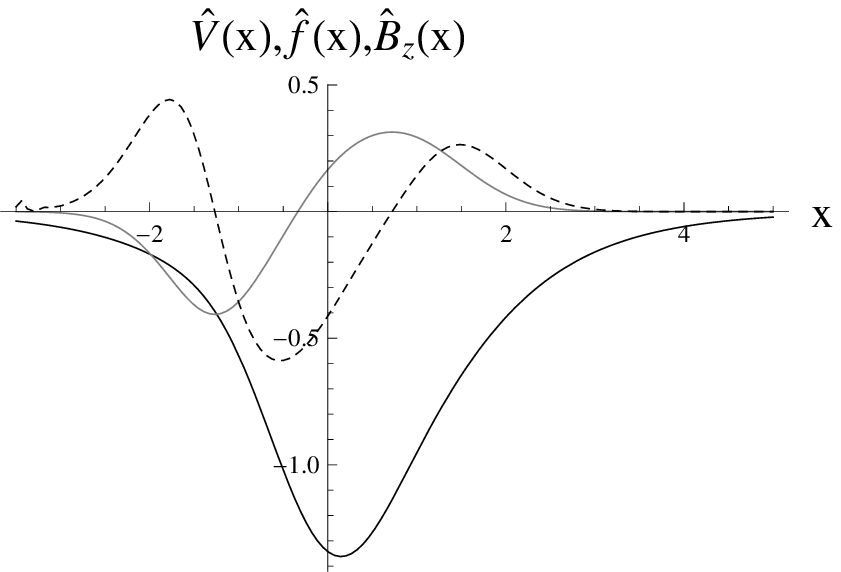,width=7.8cm}
\caption{Graphs of the transformed  scalar potential (\ref{vh32}) (black solid curves), the transformed generalized 
oscillator term (\ref{fh32}) (gray curve), and the associated $z-$component of the magnetic field (\ref{bh}) 
(dashed curve) for the mass functions $\hat{m}(x) =\exp(-x^2)$ (left plot) and 
$\hat{m}(x) = 0$ (right plot) and the settings (\ref{set3}), (\ref{h032}), (\ref{h132}), $\delta=-1$.}
\label{fig6}
\end{center}
\end{figure}

\paragraph{Second-order Darboux transformation.} Let us now apply a Darboux transformation of second order to 
our initial Dirac equation (\ref{dirac}) for the parameter settings (\ref{set3}). We need two transformation functions 
$h_0$ and $h_1$ that we define as follows
\begin{eqnarray}
h_0(x) &=& \psi_0(x)_{|k_y=-1} ~=~ \frac{\exp\left(\frac{3 x}{2}\right)}{2~\sqrt{1+\exp(2 x)}} \nonumber \\[1ex]
h_1(x) &=& \psi_0(x)_{|k_y=-2} ~=~ \frac{\exp\left(\frac{5 x}{2}\right)}{4~\sqrt{1+\exp(2 x)}}, \label{h031}
\end{eqnarray}
observe that we took $h_0$ from (\ref{h03}). Now, we insert our two transformation functions into (\ref{vj}) and calculate the 
Darboux transformation (\ref{main1}), (\ref{main2}), (\ref{main3}) for $n=2$. 
The resulting expressions, along with the present parameter settings (\ref{set3}) are then substituted 
into the term (\ref{fh}) and the scalar potential (\ref{vh}). Simplification and setting $\delta=-1$ yields
\begin{eqnarray}
\hat{V}(x) &=& -\sqrt{
\hat{m}(x)^2+\frac{20~\exp(2 x)}{\left[5+\exp(2x)\right]^2}} \label{vh31} \\[1ex]
\hat{f}(x) &=& -\frac{1}{2}+\frac{5}{5+\exp(2x)}-\frac{\hat{m}'(x)-\hat{V}'(x)}{2~\hat{m}(x)-2~\hat{V}(x)}. \label{fh31}
\end{eqnarray}
Comparison of these expressions with their first-order counterparts (\ref{vh3}) and (\ref{fh3}) shows that 
they differ merely in constants. This is due to the choice of our transformation energies as negative integers 
that render the transformation functions in elementary form. We omit to show graphs of the functions 
(\ref{vh31}) and (\ref{fh31}) because they are so similar to (\ref{vh3}) and (\ref{fh3}), respectively. Also, we do not 
display the explicit form of solutions pertaining to the transformed Dirac equation (\ref{diract}) for (\ref{vh31}) and 
(\ref{fh31}). Instead, we repeat our second-order Darboux transformation with complex conjugate 
transformation energies. More precisely, we choose our transformation functions as 
\begin{eqnarray}
h_0(x) &=& \psi_0(x)_{|k_y=-1+i} ~=~ \frac{\exp\left[\left(\frac{3}{2}-i\right)\right]x}{\sqrt{1+\exp(2 x)}} \label{h032} \\[1ex]
h_1(x) &=& \psi_0(x)_{|k_y=-1-i} ~=~ \frac{\exp\left[\left(\frac{3}{2}+i\right)\right]x}{\sqrt{1+\exp(2 x)}}. \label{h132}
\end{eqnarray}
Following our previous procedure, we substitute these two functions into (\ref{vj}), and afterwards into the 
Darboux transformation (\ref{main1}), (\ref{main2}), (\ref{main3}) for $n=2$, which in turn determines 
the term (\ref{fh}) and the  scalar potential (\ref{vh}). We find for $\delta=-1$ that
\begin{eqnarray}
\hat{V}(x) &=& -\sqrt{
\hat{m}(x)^2+\frac{260~\exp(2 x)}{\left[13+5~\exp(2x)\right]^2}
} \label{vh32} \\[1ex]
\hat{f}(x) &=& -\frac{1}{2}+\frac{13}{13+5~\exp(2x)}-\frac{\hat{m}'(x)-\hat{V}'(x)}{2~\hat{m}(x)-2~\hat{V}(x)}. 
\label{fh32}
%\hat{f}(x) &=& \frac{\left[-13+5~\exp(2 x)\right] \hat{m}(x)+\left[13+5~\exp(2 x)\right] \hat{m}'(x)}
%{2~\sqrt{260~\exp(2 x)+\left[13+5~\exp(2 x)\right]^2 \hat{m}(x)^2}}. \label{fh32}
\end{eqnarray}
The form of these functions is the same as the previous pairs (\ref{vh31}), (\ref{fh31}) and (\ref{vh3}), (\ref{fh3}). 
Examples are shown in figure \ref{fig6} for two different masses. Note that the first mass choice $\hat{m}=0$ 
makes the term (\ref{fh32}) vanish.
 \noindent

\section{Generalization to matrix potentials}\label{matrix}
In this section we shall apply Darboux transformation to a more general relativistic system, namely, 
Dirac equation in the presence of a matrix potential \cite{correa} \cite{ere} \cite{ojel} \cite{samsonov} 
and find new matrix potentials for which the Dirac equation remains solvable. 
More precisely, we consider our initial Dirac equation in the form
\begin{eqnarray}
\left\{\sigma_x \left[p_x-i~\sigma_z~f(x) \right]+\sigma_y~p_y+\sigma_z~m(x)+V(x) \right\} \Psi(x,y) &=& 0, \label{diracm}
\end{eqnarray}
where we use the same notation as in (\ref{dirac}), except that this time the potential $V=(V_{ij})$, $i,j=1,2$, 
is an arbitrary $2 \times 2$ matrix. Upon collecting terms, we can rewrite our Dirac equation as
\begin{eqnarray}
-i~\frac{\partial \Psi(x,y)}{\partial x} -i~\frac{\partial \Psi(x,y)}{\partial y}+
\left(
\begin{array}{lll}
m(x) +V_{11}(x) & i~f(x)+V_{12}(x) \\[1ex]
-i~f(x)+V_{21}(x) & -m(x)+V_{22}(x)
\end{array}
\right)
\Psi(x,y) &=& 0. \label{diracm1}
\end{eqnarray}
In the forms (\ref{diracm}) and (\ref{diracm1}), the function $f$ can be interpreted as a generalized oscillator term and 
a component of a vector potential, respectively. In the latter case the associated magnetic field is found from (\ref{b}). 
We will now approach our initial equation (\ref{diracm}) or, equivalently, its form (\ref{diracm1}) in the same way as their 
respective counterparts (\ref{dirac}) and (\ref{diracmag}) in section 3. In each step 
we can recover the latter particular case if we implement the settings $V_{11}=V_{22}=V$, 
$V_{12}=V_{21}=0$. Let us now substitute (\ref{psi}) into (\ref{diracm1}), resulting in the component equations
\begin{eqnarray}
-i~\Psi_2'(x)+\left[
-i~k_y+i~f(x)+V_{12}(x)\right]\Psi_2(x)+
\left[ m(x)+V_{11}(x)\right]\Psi_1(x) &=& 0 \label{sys1m} \\[1ex]
-i~\Psi_1'(x)+\left[
i~k_y-i~f(x)+V_{21}(x)\right]\Psi_1(x)+
\left[- m(x)+V_{22}(x)\right]\Psi_2(x) &=& 0. \label{sys2m}
\end{eqnarray}
We solve the second component equation with respect to $\Psi_2$. This yields
\begin{eqnarray}
\Psi_2(x) &=& \frac{[i~f(x)-i~k_y-V_{21}(x)]~\Psi_1(x)+i~\Psi_1'(x)}{V(x)-m(x)}. \label{psi2m}
\end{eqnarray}
The remaining component (\ref{sys1m}) can be rewritten by redefining $\Psi_1$ as
\begin{eqnarray}
\Psi_1(x) &=& \exp\left[-\frac{i}{2} \int\limits^x V_{12}(t)+V_{21}(t)~dt\right]
\sqrt{m(x)-V_{22}(x)}~\psi_0(x). \nonumber
\end{eqnarray}
Upon implementing this definition in (\ref{sys1m}), we obtain the following Schr\"odinger-type equation for the function $\psi_0$
\begin{eqnarray}
\psi_0''(x)- \left[k_y^2+k_y~X_0(x)+Y_0(x)\right] \psi_0(x) &=& 0, \label{ssem}
\end{eqnarray}
where the potential term $X_0$ is given explicitly by
\begin{eqnarray}
X_0(x) &=& -2~f(x) +i~[V_{12}(x)-V_{21}(x)]- \frac{m'(x)-V_{22}'(x)}{m(x)-V_{22}(x)}. \label{x0m} 
\end{eqnarray}
Since the remaining potential term $Y_0$ has a very long and involved form, we omit to state it explicitly here. Before we continue, 
let us briefly comment on a simplification of our Schr\"odinger-type equation (\ref{ssem}) that occurs for $X_0=0$. 
Similar to the setting (\ref{fsim}) worked out in the previous section, we fix our term $f$ to 
be given as
\begin{eqnarray}
f(x) &=& -\frac{m'(x)-V_{22}'(x)}{2~m(x)-2~V_{22}(x)}+\frac{i}{2}~\left[V_{12}(x)-V_{21}(x)\right]. \label{fsimm}
\end{eqnarray}
This setting forces $X_0=0$ and furthermore renders our equation (\ref{ssem}) in the compact form
\begin{eqnarray}
\psi_0''(x)+ \left\{-k_y^2+\left[m(x)+V_{11}(x) \right] \left[ m(x)-V_{22}(x) \right] \right\} \psi_0(x) &=& 0. \label{schrm}
\end{eqnarray}
We observe that this generalization of (\ref{schr}) resembles a conventional Schr\"odinger equation, where 
$-k_y^2$ plays the role of the stationary energy. Now let us return to our Darboux transformation. After applying 
the latter transformation (\ref{main1}), (\ref{main2}), (\ref{main3}) to equation (\ref{ssem}), we obtain its transformed 
counterpart as
\begin{eqnarray}
\psi_n''(x)- \left[k_y^2+k_y~X_n(x)+Y_n(x)\right] \psi_n(x) &=& 0. \label{ssetm}
\end{eqnarray}
Our next step consists in matching the form of the transformed potential terms with their initial partners. Our goal 
is to transfer (\ref{ssetm}) to our transformed Dirac equation
\begin{eqnarray}
\left\{\sigma_x \left[p_x-i~\sigma_z~\hat{f}(x) \right]+\sigma_y~p_y+\sigma_z~\hat{m}(x)+\hat{V}(x) \right\} \hat{\Psi}(x,y) 
&=& 0, \label{diractm}
\end{eqnarray}
where we adopt the notation from (\ref{diract}) except for the transformed potential $\hat{V}=(\hat{V}_{ij})$, $i,j=1,2$, 
representing a matrix rather than a function. Similar to (\ref{xncon}), our matching condition for $X_n$ reads
\begin{eqnarray}
-2~f(x) +i~[V_{12}(x)-V_{21}(x)]- \frac{m'(x)-V_{22}'(x)}{m(x)-V_{22}(x)} +\Delta X_n(x)&=& \nonumber \\[1ex]
& & \hspace{-6cm} =~
 -2~\hat{f}(x) +i~[\hat{V}_{12}(x)-\hat{V}_{21}(x)]- \frac{\hat{m}'(x)-\hat{V}_{22}'(x)}{\hat{m}(x)-\hat{V}_{22}(x)}. 
\label{xnconm}
\end{eqnarray}
Furthermore, note that we implemented the abbreviation $\Delta X_n$ from (\ref{deltas}). We can solve 
our condition (\ref{xnconm}) with respect to the term $\hat{f}$ as
\begin{eqnarray}
\hat{f}(x) &=& f(x)+\frac{i}{2} \left[\hat{V}_{12}(x)-\hat{V}_{21}(x)\right]
-\frac{i}{2} \left[V_{12}(x)-V_{21}(x)\right]+\frac{m'(x)-V_{22}'(x)}{2~m(x)-2~V_{22}(x)}
+ \nonumber \\[1ex]
&+&\frac{\left[\hat{V}_{22}(x)-\hat{m}(x)\right] \Delta X_n(x)-\hat{m}'(x)+\hat{V}_{22}'(x)}{2~\hat{m}(x)-2~\hat{V}_{22}(x)}. 
\label{fhm}
\end{eqnarray}
Next we must solve the remaining condition pertaining to the potential term $Y_n$ in (\ref{ssetm}). Since we avoid to 
state $Y_n$ explicitly, we give the latter condition in abbreviated form as
\begin{eqnarray}
Y_n(x) &=& Y_0(x)_{\mid f \rightarrow \hat{f}, m \rightarrow \hat{m}, V_{ij} \rightarrow \hat{V}_{ij}}. \label{ynconm}
\end{eqnarray}
We point out that the function $\hat{f}$ is given by (\ref{fhm}). Upon insertion of this function we can solve condition 
(\ref{ynconm}) with respect to $\hat{m}$, $\hat{V}_{11}$, and $\hat{V}_{22}$. We cannot use the off-diagonal potential 
matrix entries $\hat{V}_{12}$ or $\hat{V}_{21}$ to solve (\ref{ynconm}) because they do not occur in our condition. 
Let us now state the three solutions mentioned above. When solving for the mass function $\hat{m}$, we obtain
\begin{eqnarray}
\hat{m}(x) &=& \frac{\hat{V}_{11}(x)~V_{22}(x)-V_{22}(x)~\hat{V}_{22}(x)-\hat{V}_{11}(x)~m(x)+\hat{V}_{22}(x)~m(x)}
{2~m(x)-2~V_{22}(x)}+ \nonumber \\[1ex]
&+&
\frac{1}{2~m(x)-2~V_{22}(x)}~\Bigg\{
\left[m(x)-V_{22}(x) \right] \Bigg\{4~m(x)^3+4~m(x)^2~\Bigg[V_{11}(x)-2~V_{22}(x)\Bigg]- \nonumber \\[1ex]
&-&
\hat{V}_{11}(x)^2~V_{22}(x)
+4~V_{11}(x)~V_{22}(x)^2-2~\hat{V}_{11}(x)~\hat{V}_{22}(x)-V_{22}(x)~\hat{V}_{22}(x)^2-
\nonumber \\[1ex]
&-&
4~f(x)~V_{22}(x)~\Delta X_n(x)
+2~i~V_{12}(x)~V_{22}(x)~\Delta X_n(x)-2~i~V_{21}(x)~V_{22}(x)~\Delta X_n(x)+
\nonumber \\[1ex]
&+&
V_{22}(x)~\Delta X_n(x)^2
-4~V_{22}(x)~\Delta Y_n(x)+2~\Delta X_n(x)~m'(x)-2~\Delta X_n(x)~V_{22}'(x)+
\nonumber \\[1ex]
&+&
m(x) \Bigg[-8~V_{11}(x)~V_{22}(x)+4~V_{22}(x)^2+\Bigg(\hat{V}_{11}(x)+\hat{V}_{22}(x)\Bigg)^2+
\Delta X_n(x) \times 
\nonumber \\[1ex]
&\times&
\Bigg(4~f(x)-2~i~V_{21}(x)-\Delta X_n(x)\Bigg)+4~\Delta Y_n(x)-2~\Delta X_n'(x)\Bigg]+
\nonumber \\[1ex]
&+&
2~V_{22}(x)~\Delta X_n'(x) \Bigg\}
\Bigg\}^{\frac{1}{2}}. \nonumber 
\end{eqnarray}
Let us now solve our condition (\ref{ynconm}) with respect to the transformed matrix potential entry $\hat{V}_{11}$. 
Our result reads
\begin{eqnarray}
\hat{V}_{11}(x) &=& \frac{1}{4~[m(x)-V_{22}(x)]~[\hat{m}(x)-\hat{V}_{22}(x)]}~
\Bigg\{4~m(x)^3+4~m(x)^2~V_{11}(x)-
\nonumber \\[1ex]
&-&
8~m(x)^2~V_{22}(x)+4~\hat{m}(x)^2~V_{22}(x)+4~V_{11}(x)~V_{22}(x)^2-
\nonumber \\[1ex]
&-&
4~\hat{m}(x)~V_{22}(x)~\hat{V}_{22}(x)-
4~f(x)~V_{22}(x)~\Delta X_n(x)+
\nonumber \\[1ex]
&+&
2~i~V_{12}(x)~V_{22}(x)~\Delta X_n(x)
-2~i~V_{21}(x)~V_{22}(x)~\Delta X_n(x)+V_{22}(x)~\Delta X_n(x)^2-
\nonumber \\[1ex]
&-&
4~V_{22}(x)~\Delta Y_n(x)+
2~\Delta X_n(x)~m'(x)-2~\Delta X_n(x)~V_{22}'(x)+
2~V_{22}(x)~\Delta X_n'(x)-
\nonumber \\[1ex]
&-&
8~m(x)~V_{11}(x)~V_{22}(x)+4~m(x)~V_{22}(x)^2-4~m(x)~\hat{m}(x)^2+
\nonumber \\[1ex]
&+&
4~m(x)~\hat{m}(x)~\hat{V}_{22}(x)+4~m(x)~f(x)~\Delta X_n(x)-
\nonumber \\[1ex]
&-&
2~i~m(x)~\Delta X_n(x)~[V_{12}(x)-V_{21}(x)]
-m(x)~\Delta X_n(x)^2+4~m(x)~\Delta Y_n(x)-
\nonumber \\[1ex]
&-&
2~m(x)~\Delta X_n'(x)
\Bigg\} \label{vh11sol}
\end{eqnarray}
As mentioned above, we can also solve condition (\ref{ynconm}) for the transformed matrix potential entry 
$\hat{V}_{22}$. However, the solution is very similar to (\ref{vh11sol}) in the following sense: if we replace 
$\hat{V}_{22}$ in (\ref{vh11sol}) by $-\hat{V}_{11}$, then we obtain the solution of (\ref{ynconm}) with respect to 
$\hat{V}_{11}$. For this reason we will not state its explicit form here.

\begin{figure}[h]
\begin{center}
\epsfig{file=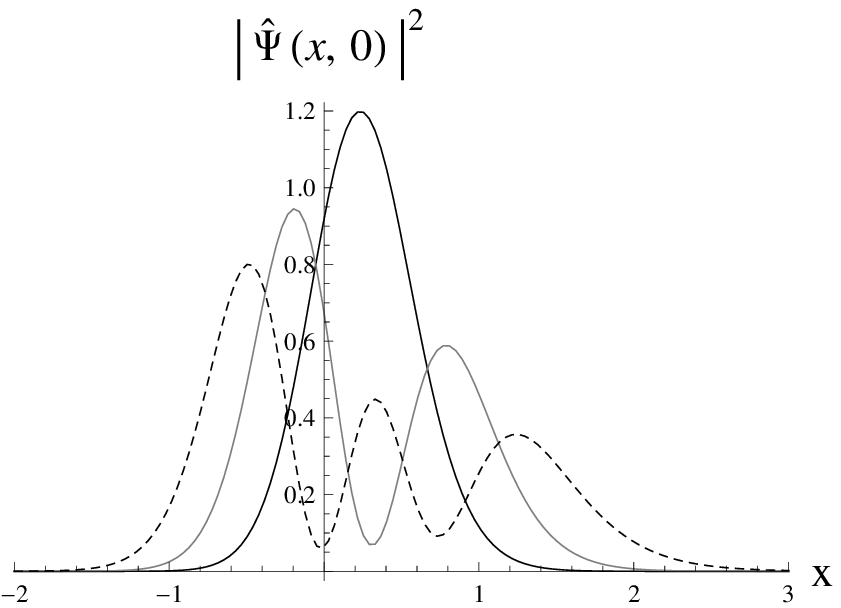,width=7.8cm}
\epsfig{file=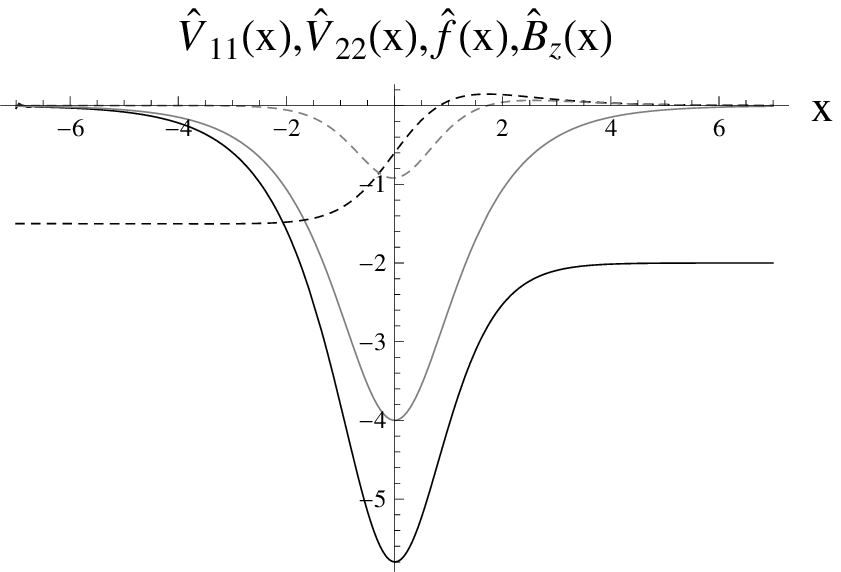,width=7.8cm}
\caption{Left plot: graphs of the normalized probability densities $|\hat{\Psi}(x,0)|^2$ associated with the solution 
(\ref{psih}) of our transformed Dirac equation (\ref{diractm}) for the settings (\ref{setm}), (\ref{psi01}), (\ref{h01}). 
Parameter values are $k_y=4$ (black solid curve), 
$k_y=3$ (gray curve), and $k_y=2$ (dashed curve). Right plot: graphs of the entries $\hat{V}_{11}$ (black solid curve), 
$\hat{V}_{22}$ (gray curve) pertaining to the matrix (\ref{vhmx}), the oscillator term (\ref{fhmx}) (black dashed curve), 
and the $z-$component of the magnetic field (\ref{magm}) (gray dashed curve).}
\label{fig8}
\end{center}
\end{figure}

\paragraph{First-order Darboux transformation.} In this paragraph we will demonstrate how our Darboux transformation 
works in practice if the potential in our Dirac equation (\ref{diracm}) is not a multiple of the identity matrix. To this end, 
let us first specify our initial parameter settings.
\begin{eqnarray}
f(x) ~=~\frac{1}{2}~\tanh(x) \qquad \qquad V(x) ~=~ \sqrt{30}~\mbox{sech}(x)~I_2 \qquad \qquad m(x) ~=~ 0. \label{setm}
\end{eqnarray}
We observe that these settings are the same as (\ref{set1}), note that our notation has changed due to $V$ now being an 
actual matrix. Consequently, our initial Dirac equation (\ref{diracm}) for the settings (\ref{setm}) is the same as its 
former counterpart (\ref{dirac}) with (\ref{set1}). We can therefore use the Schr\"odinger solution (\ref{psi01}) and the 
transformation function (\ref{h01}) for our Darboux transformation. We first substitute the latter two function along with 
(\ref{vj}) into (\ref{main1}), (\ref{main2}), (\ref{main3}) for $n=1$. In the subsequent step we insert the results into the 
transformed term (\ref{fhm}) . Simplification leads to 
the findings
\begin{eqnarray}
\hat{f}(x) &=& \frac{1}{2~\hat{m}(x)-2~\hat{V}_{22}(x)} ~\Bigg\{
\hat{m}(x)~\Bigg[-1+\tanh(x)+i~\hat{V}_{12}(x)-i~\hat{V}_{21}(x)\Bigg]+\hat{V}_{22}(x) - \nonumber \\[1ex]
&-& \Bigg[\tanh(x)~\hat{V}_{22}(x)+i~\hat{V}_{12}(x)~\hat{V}_{22}(x)-i~\hat{V}_{21}(x)~\hat{V}_{22}(x)\Bigg]
-\hat{m}'(x)+\hat{V}_{22}'(x)
\Bigg\}. \label{fhm1}
\end{eqnarray}
In a similar way we can determine the transformed potential matrix $\hat{V}$ by substitution of our current parameters 
into (\ref{vh11sol}). We obtain
\begin{eqnarray}
\hat{V}(x) &=&
\left(
\begin{array}{cc}
{\displaystyle{-\hat{m}(x)+\frac{24~\mbox{sech}(x)^2}{\hat{V}_{22}(x)-\hat{m}(x)}}} & \hat{V}_{12}(x) \\[3ex]
\hat{V}_{21}(x) & \hat{V}_{22}(x)
\end{array}
\right). \label{vhm1}
\end{eqnarray}
We observe that the transformed mass function and three entries of the transformed potential matrix remain 
undetermined, allowing to generate a wide variety of Dirac equations (\ref{diractm}), along with its associated solutions. 
Let us now state an example by introducing the settings
\begin{eqnarray}
\hat{m}(x)&=&1+\tanh(x) \qquad \qquad \hat{V}_{22}(x)~=~-4~\mbox{sech}(x) \qquad \qquad 
\hat{V}_{12}(x)~=~\hat{V}_{21}(x)~=~0. \label{setadd}
\end{eqnarray}
If we plug these settings into the term (\ref{fhm1}), we obtain its explicit form
\begin{eqnarray}
\hat{f}(x) &=& -\frac{\mbox{sech}(x) \left[2+\mbox{sech}(x)-4~\tanh(x)\right]}{1+4~\mbox{sech}(x)+\tanh(x)}. 
\label{fhmx}
\end{eqnarray}
The magnetic field (\ref{bh}) generated by this function can be calculated as
\begin{eqnarray}
\hat{B}(x) &=& \left(0,0,\frac{2~\exp(x)}{[4+\exp(x)]^2}-\frac{4}{\exp(-x)+\exp(x)]^2}\right)^T. \label{magm}
\end{eqnarray}
The $z-$component of the magnetic field is visualized in the right part of figure \ref{fig8}. 
The transformed matrix potential is found by inserting our current settings (\ref{setm}) and (\ref{setadd}) 
into (\ref{vhm1}). The resulting potential has the form
\begin{eqnarray}
\hat{V}(x) &=& \left(
\begin{array}{cc}
{\displaystyle{-\frac{\left[25+4~\mbox{sech}(x)-22~\tanh(x)\right] \left[1+\tanh(x)\right]}{1+4~\mbox{sech}(x)+\tanh(x)}}} 
& 0 \\[3ex]
0 & -4~\mbox{sech}(x)
\end{array}
\right). \label{vhmx}
\end{eqnarray}
Both the term (\ref{fhmx}) and the non-vanishing potential components from (\ref{vhmx}) are shown 
in the right part of figure \ref{fig8}. Since the explicit form of the associated solutions to the transformed Dirac 
equation (\ref{diractm}) is very long, we omit to show it here. Instead, we visualize the corresponding probability 
densities in the left part of figure \ref{fig8}. 
 \noindent

\section{Concluding remarks} \label{con}
The Darboux transformation presented in this work is applicable to Dirac equations at zero energy with magnetic field, 
position-dependent mass and matrix potential, including the special cases of vanishing mass and scalar potential. 
Instead of being coupled to a magnetic field, our systems can also be interpreted as generalized Dirac oscillators due to a 
one-to-one correspondence between the two scenarios. A particular feature of our approach is that 
the position-dependent mass in the Darboux-transformed Dirac equation remains undetermined and can be chosen 
arbitrarily. This property is useful for example when comparing exactly-solvable massless systems 
(such as in Dirac materials) to their massive counterparts. It should be pointed out that the algorithm 
summarized in section 2 is not equivalent to the conventional Darboux transformation,also referred to as SUSY formalism. 
As such, the results we obtain here cannot be found through application of the latter formalism. The extension of 
the present method to more general systems like bilayer graphene is subject of future research.

\end{sloppypar}

\end{document}